\documentclass[aps,pra,superscriptaddress,twocolumn]{revtex4-1}
\usepackage{xcolor}
\usepackage{natbib}
\usepackage{amsmath}
\usepackage{amssymb}
\usepackage{amsthm}
\usepackage{nicefrac}
\usepackage{graphicx}
\usepackage{bm}
\usepackage{url}
\usepackage{physics}

\renewcommand{\vec}[1]{\bm{\mathrm{#1}}}
\renewcommand{\cos}[1]{\text{cos}\left(#1\right)}
\renewcommand{\sin}[1]{\text{sin}\left(#1\right)}

\newcommand{\vecop}[1]{\hat{\mathbf{#1}}}

\newcommand{\MatrixVariable}[1]{\ensuremath{\overline{\overline{#1}}}}%

\begin{document}

\title{Trace Expressions and Associated Limits for Non-Equilibrium Casimir Torque}

\author{Benjamin Strekha}
\affiliation{Department of Electrical and Computer Engineering, Princeton University, Princeton, New Jersey 08544, USA}

\author{Sean Molesky}
\affiliation{Department of Engineering Physics, Polytechnique Montréal, Montréal, Québec H3T 1J4, Canada}

\author{Pengning Chao}
\affiliation{Department of Electrical and Computer Engineering, Princeton University, Princeton, New Jersey 08544, USA}

\author{Matthias Kr\"{u}ger}
\affiliation{Institute for Theoretical Physics, Georg-August-Universit\"{a}t G\"{o}ttingen, 37073 G\"{o}ttingen, Germany}

\author{Alejandro W. Rodriguez}
\affiliation{Department of Electrical and Computer Engineering, Princeton University, Princeton, New Jersey 08544, USA}


\begin{abstract}
We exploit fluctuational electrodynamics to present trace expressions
for the torque experienced by arbitrary objects in a passive,
non-absorbing, rotationally invariant background environment.
Specializing to a single object, this formalism, together with
recently developed techniques for calculating bounds via Lagrange
duality, is then used to derive limits on the maximum Casimir torque
that a single object with an isotropic electric susceptibility can
experience when out of equilibrium with its surrounding
environment. The maximum torque achievable at any wavelength is shown
to scale in proportion to body volumes in both subwavelength
(quasistatics) and macroscopic (ray optics) settings, and come within
an order of magnitude of achievable torques on topology optimized
bodies. Finally, we discuss how to extend the formalism to multiple
bodies, deriving expressions for the torque experienced by two
subwavelength particles in proximity to one another.
\end{abstract}

\maketitle

Over the past decades, much effort has been devoted to understanding
fluctuation phenomena in structured
media~\cite{biehs2021near,woods2016materials}. For example, in recent
years Casimir forces have been considered in a variety of systems out
of thermal equilibrium, including
planar slabs~\cite{antezza2008casimirtwoplates},
spheres~\cite{kruger2011spheresandplates,kruger2011spheresandplates,kruger_trace_formulae_for_nonequilibrium,muller2016anisotropic},
cylinders~\cite{golyk2012casimircylinders}, and
gratings~\cite{noto2014casimirgratings}.  Whether through surface
texturing or by enforcing far out of equilibrium conditions, the
Casimir force can be made to exhibit a wide range of power
laws~\cite{antezza2006casimir}, lead to unstable and stable
equilibria~\cite{kruger2011spheresandplates}, become
repulsive~\cite{antezza2008casimirtwoplates,kruger2011spheresandplates,kruger2011spheresandplates},
and lead to
self-propulsion~\cite{kruger2011spheresandplates,muller2016anisotropic}.
In anisotropic media or systems exhibiting chirality, thermal
fluctuations can also cause objects to exchange net angular momentum
with their environments or other nearby objects, resulting in a net
torque~\cite{kats1971nonisotropic,chinmay2021forcetorque,parsegian1972dielectric,guo2021single,gao2021thermal},
a prediction that was recently verified in
experiments~\cite{somers2018measurement}. As interest in mechanical
devices of increasingly smaller scales continues to grow, so too is
the ability to exploit fluctuation phenomena such as laser shot noise
and the Casimir effect to actuate nanocale
rotors~\cite{ding2022universal,van2021sub,stickler2021quantum}.

In this paper, we exploit the mathematical framework of fluctuational
electrodynamics~\cite{rytov1989principles,otey2014fluctuational} and
scattering theory~\cite{rahi2009scattering} to rigorously derive trace
expressions for the thermal Casimir torque experienced by a set of
objects out of equilibrium with themselves or their environment. Based
on the dyadic Green's function $\mathbb{G}_{0}$ of a rotationally
invariant background environment and the scattering $\mathbb{T}$
operators of each object in isolation, these expressions, valid also
for arbitrary anisotropic bodies, are extensions of analogous and
recently derived power and force
quantities~\cite{kruger_trace_formulae_for_nonequilibrium}. Special
attention is given to the case of a single body embedded in a
background environment as well as two-body scenarios, generalizing
recent expressions for the torque on dipolar
particles~\cite{manjavacas2010vacuum,manjavacas2010thermal,guo2021single,chinmay2021forcetorque}. Furthermore,
employing Lagrange duality in the case where the single body is
composed of an isotropic electric susceptibility, we present upper
bounds on the frequency contributions to the non-equilibrium Casimir
torque possible for arbitrarily structured objects confined within a
bounding sphere. These bounds show that the maximum torque experienced
by a body scales like the volume of the object in both the small
particle (quasistatics) and large-body (ray optics) limits, a feature
unique to torque as both heat transfer and forces are known to scale
like area in the large-size
limit~\cite{T_operator_bounds_angle_integrated,woods2016materials}.
Finally, our expressions are valid for arbitrary geometries and take
simple forms in the limit of dipolar particles, which we illustrate by
deriving expressions involving torque between two subwavelength bodies
out of equilibrium and in the vicinity of one another. For clarity and
conciseness, the main text focuses on fundamental equations and
results, leaving detailed derivations and technical discussions to the
appendix; interested readers are encouraged to consult the appendix
for additional insights.

\section{General Force and Torque Formulas}

The net force and torque on a body resulting from a set of prescribed
electromagnetic fields $\vec{E}$ and $\vec{B}$ acting on it can be
derived from the Lorentz force law~\footnote{The expressions are in
  the Fourier frequency space and there is only one frequency integral
  here in order to simplify the expressions, since ultimately we will
  perform an ensemble average where the different frequency components
  are uncorrelated, according to the fluctuation-dissipation
  theorem.}, and are given by (using Einstein convention):
\begin{align}
    \mathbf{F} &= \int_{V} d^{3}\mathbf{r}\int_{-\infty}^{\infty} d\omega \left[ \rho^{*}(\omega, \mathbf{r}) \mathbf{E}(\omega, \mathbf{r}) + \mathbf{J}^{*}(\omega, \mathbf{r})\times\mathbf{B}(\omega,\mathbf{r}) \right] \\
    &= \int_{V,\omega} \left[ \frac{i}{\omega}(\nabla\cdot \mathbf{J}^{*})\mathbf{E} - \frac{i}{\omega}\mathbf{J}^{*}\times(\nabla \times \mathbf{E})\right], \\
    &= \int_{V,\omega}\frac{i}{\omega}\left[ (\nabla\cdot \mathbf{J}^{*})\mathbf{E} - \mathbf{J}^{*}\cdot \nabla\mathbf{E}(\omega,\mathbf{r}) + (\mathbf{J}^{*}\cdot\nabla)\mathbf{E}\right], \\
    &= \int_{V,\omega} \frac{i}{\omega} \left[ \frac{\partial J_{j}^{*}}{\partial r_{j}} E_{k} \mathbf{e}_{k} - J_{j}^{*}\frac{\partial E_{j}}{\partial r_{k}}  \mathbf{e}_{k} + J_{j}^{*}\frac{\partial E_{k}}{\partial r_{j}} \mathbf{e}_{k} \right], \\
    &= \int_{V} d^{3}\mathbf{r}\int_{-\infty}^{\infty} d\omega\frac{1}{\hbar\omega} \left[  J_{a}^{*} \vecop{p} E_{a} + i\hbar\frac{\partial}{\partial r_{j}}(J_{j}^{*}E_{k}) \mathbf{e}_{k} \right]
    \label{eq:compactforceintegral}
\end{align}

\begin{align}
    \boldsymbol{\tau}
    &= \int_{V} d^{3}\mathbf{r}\int_{-\infty}^{\infty} d\omega \mathbf{r}\times \left[ \rho^{*}(\omega, \mathbf{r}) \mathbf{E}(\omega, \mathbf{r}) + \mathbf{J}^{*}(\omega, \mathbf{r})\times\mathbf{B}(\omega,\mathbf{r}) \right] \\
    &= \int_{V,\omega}
    \frac{i}{\omega}\mathbf{r}\times\left[ (\nabla\cdot \mathbf{J}^{*})\mathbf{E} - \mathbf{J}^{*}\times(\nabla \times \mathbf{E})\right]\\
    &= \int_{V,\omega}
    \frac{i}{\omega}\mathbf{e}_{m}\epsilon_{mjk}[-r_{j}J_{a}^{*}(\partial_{k}E_{a}) + \partial_{a}(r_{j}J_{a}^{*}E_{k}) - J_{j}^{*}E_{k}] \\
    &= \int_{V} d^{3}\mathbf{r}\int_{-\infty}^{\infty} d\omega
    \frac{1}{\hbar\omega}[ J_{a}^{*} \vecop{L} E_{a} +  \mathbf{J}^{*}\vecop{S}\mathbf{E} +\nonumber \\
    &\qquad \qquad \qquad \qquad \ \ \ \ 
    i\hbar\mathbf{e}_{m}\epsilon_{mjk}\partial_{a}(r_{j}J_{a}^{*}E_{k})]
    \label{eq:compacttorqueintegral}
\end{align}
where $\hat{\mathbf{L}} = \mathbf{r}\times\hat{\mathbf{p}} = -i\hbar\mathbf{r}\times\nabla$ and
\begin{align}
    \vecop{S} = -i\hbar
    \Bigg\{ 
    \begin{bmatrix}
        0 & 0 & 0 \\
        0 & 0 & 1 \\
        0 & -1 & 0 
    \end{bmatrix},
    \begin{bmatrix}
        0 & 0 & -1 \\
        0 & 0 & 0 \\
        1 & 0 & 0 
    \end{bmatrix},
    \begin{bmatrix}
        0 & 1 & 0 \\
        -1 & 0 & 0 \\
        0 & 0 & 0 
    \end{bmatrix}
    \Bigg\},
\end{align}
are the orbital and spin angular momentum operators,
respectively~\cite{khersonskii1988quantum}, defined in the Cartesian
basis and compactly summarized by $(\hat{S}_{a})_{bc} = -i\hbar
\epsilon_{abc}$. In deriving the expressions above, we made use of the
continuity equation $i\omega\rho(\omega, \mathbf{r}) =
\nabla\cdot\mathbf{J}(\omega, \mathbf{r})$, Faraday's law
$\mathbf{B}(\omega,\mathbf{r}) = \frac{1}{i\omega}\nabla\times
\mathbf{E}(\omega,\mathbf{r})$, and the following algebraic
identities:
\begin{align*}
    &\mathbf{r}\times(\mathbf{J}^{*}(\omega, \mathbf{r})\times(\nabla \times \mathbf{E}(\omega, \mathbf{r}))) \\ &= 
    \mathbf{e}_{i}\epsilon_{ijk}r_{j}(\mathbf{J}^{*}(\omega, \mathbf{r})\times(\nabla \times \mathbf{E}(\omega, \mathbf{r})))_{k} \\
    &=
    \mathbf{e}_{i}\epsilon_{ijk}r_{j}
    (
    J_{a}^{*}(\partial_{k}E_{a}) - J_{a}^{*}(\partial_{a}E_{k})
    )\\
    &=
    \mathbf{e}_{i}\epsilon_{ijk}
    (
    r_{j}J_{a}^{*}(\partial_{k}E_{a}) - \partial_{a}(r_{j}J_{a}^{*}E_{k}) \\ &\hspace{1.2in}+ \delta_{aj}J_{a}^{*}E_{k} + r_{j}(\partial_{a}J_{a}^{*})E_{k}
    )
\end{align*}    
and $\mathbf{r}\times(\nabla\cdot
\mathbf{J}^{*}(\omega,\mathbf{r}))\mathbf{E}(\omega, \mathbf{r}) =
\mathbf{e}_{i}\epsilon_{ijk}r_{j}E_{k}(\partial_{a}J_{a}^{*})$.

Notably, the total derivative terms $\sim
\partial_{a}(r_{j}J_{a}^{*}E_{k})$ above vanish in scenarios in which
there are no net currents just outside the body. In fact,
Eq.~\eqref{eq:compactforceintegral} minus the total derivative terms
has been used as the starting point for deriving trace expressions for
the Casimir
force~\cite{kruger_trace_formulae_for_nonequilibrium,muller2016anisotropic,kruger2011spheresandplates,gelbwaser2021near}.
In considering torque, one might naively though incorrectly insert
$\mathbf{r}\times$ into prior trace expressions for
forces~\cite{kruger_trace_formulae_for_nonequilibrium,muller2016anisotropic,kruger2011spheresandplates},
introducing terms of the form $\mathbf{r}\times\nabla$ and thus
leading to quantities proportional to the orbital angular momentum
operator $\hat{\mathbf{L}} = \mathbf{r}\times\hat{\mathbf{p}} =
-i\hbar\mathbf{r}\times\nabla.$ Specifically, while the
$J_{a}^{*}\vecop{L} E_{a}$ term above would follow upon inserting
$\mathbf{r}\times$ into the force expression of
Eq.~\eqref{eq:compactforceintegral}, such naive manipulation would
miss the additional term $\mathbf{J}^{*}\vecop{S}\mathbf{E} =
-i\hbar\mathbf{J}^{*}\times\mathbf{E}$ present in
Eq.~\eqref{eq:compacttorqueintegral}. The presence of this last term should be expected on
physical grounds: a photon is a spin-1 particle, and the torque
exerted by a vector field does not just depend on angular derivatives
(``orbital'' contributions), but also on the mixing of different
vector components (``spin'' contributions).

\section{Casimir Torque on a Single Body}
Starting from the above general expression, one can derive a
corresponding expression for the Casimir torque on a collection of
objects, the origin of which are thermal fluctuations of currents and
fields in matter and throughout space. The relation quantifying the
statistical thermodynamics of matter and resulting charge fluctuations
is known as the fluctuation-dissipation theorem (FDT), and takes the
form~\cite{novotny2012principles,kruger_nonequilbrium_fluctuations_review}
\begin{multline}
\langle  J_{i}(\mathbf{x}, \omega) J_{j}^{*}(\mathbf{x}', \omega ')
\rangle_{T}\\ = \frac{\omega\epsilon_{0}}{2\pi} \coth\left( \frac{\hbar
  \omega}{2k_{B}T} \right)
\bm{\chi}_{ij}^{\mathsf{A}}(\mathbf{x},\mathbf{x}';\omega) \delta(\omega - \omega'), 
\label{eq:FDT}
\end{multline}
with $\langle\cdots\rangle_{T}$ denoting an equilibrium thermal
average of the electric current sources in a
medium of general electric susceptibility $\bm{\chi}$ held at a
temperature $T$. The superscript $\mathsf{A}$ on an operator $\Theta$
denotes its asymmetric part, $\Theta^{\mathsf{A}} \equiv
\frac{1}{2i}(\Theta - \Theta^{\dagger}),$ where $\dagger$ denotes
conjugate transpose. In our notation,
$\Theta_{ab}^{\dagger}(\mathbf{x}, \mathbf{y}) =
\Theta_{ba}(\mathbf{y}, \mathbf{x})^{*}$, treating the vector
component and spatial coordinate as an index pair. For systems in thermal
equilibrium, the current--current correlations along with Maxwell's
equations can be used to derive corresponding field--field
correlations, $\mathbb{C}_{ij}^{eq}(T, \omega,\omega'; \mathbf{r},
\mathbf{r}') \equiv \langle E_{i}(\mathbf{r},
\omega)E_{j}^{*}(\mathbf{r}', \omega ') \rangle_{T} =
\frac{\hbar\omega^{2}}{2\pi c^{2}\epsilon_{0}} \coth\left(
\frac{\hbar\omega}{2k_{B}T} \right) \delta(\omega - \omega')
\mathbb{G}^{\mathsf{A}}_{ij}(\omega; \mathbf{r}, \mathbf{r}')$, in
terms of the Green's function of the system $\mathbb{G}$, defined by
$\left[ \nabla\times\nabla\times - \mathbb{V} -
  \frac{\omega^{2}}{c^{2}}\mathbb{I}\right]\mathbb{G}(\mathbf{r},
\mathbf{r}') = \mathbb{I}\delta^{(3)}(\mathbf{r} - \mathbf{r}')$ where
$\mathbb{V} = \frac{\omega^{2}}{c^{2}}(\mathbb{\epsilon} - \mathbb{I})
+ \nabla\times(\mathbb{I} - \mathbb{\mu}^{-1})\nabla\times$ is the
potential or generalized susceptibility introduced by the
objects~\cite{kruger_nonequilbrium_fluctuations_review}.  In a
nonequilibrium stationary state, each object is assumed to be at local
equilibrium, such that the current fluctuations within each object
satisfy the FDT at the appropriate local temperature. The details of
the use of FDT and local equilibrium properties with the scattering
equations have been described
before~\cite{kruger_nonequilbrium_fluctuations_review,kruger_trace_formulae_for_nonequilibrium}
and laid out in the appendix. Detailed derivations and use of similar
principles to derive force expressions can be found in
Refs.~\cite{kruger_trace_formulae_for_nonequilibrium,muller2016anisotropic}. Here,
we restrict our attention to torque.


As a concrete example, we consider the Casimir torque on an isolated
body out of equilibrium with its surroundings. To begin with, we
use the linear response relation between the set of sources in the
body and resulting fields, $\mathbf{E} =
i\mu_{0}\omega\mathbb{G}_{0}\mathbf{J}$, to rewrite the thermally
averaged torque in terms of the field--field correlation Dyadic:
\begin{align}
    \boldsymbol{\tau} &= \textrm{Re}\int_{V_{body}} d^{3}\mathbf{r}\int_{-\infty}^{\infty} d\omega
    \frac{1}{\hbar\omega}[J_{a}^{*} \vecop{L} E_{a} +  \mathbf{J}^{*}\vecop{S}\mathbf{E}]  \\
    &= -\textrm{Im}\int_{V_{body}} d^{3}\mathbf{r}\int_{-\infty}^{\infty} d\omega
    \frac{1}{\hbar\omega^{2}\mu_{0}}[ (\mathbb{G}_{0}^{-1}\mathbf{E})_{a}^{*} \vecop{L} E_{a} + \nonumber \\ &\qquad\qquad\qquad\qquad
    \qquad\qquad\qquad(\mathbb{G}_{0}^{-1}\mathbf{E})^{*}\vecop{S}\mathbf{E}] \label{eq:torquetrace1} \\
    &= -\textrm{Im}\int_{-\infty}^{\infty} d\omega
    \frac{1}{\hbar\omega^{2}\mu_{0}} \text{Tr}|_{V_{body}}[\vecop{J} \mathbb{C}\mathbb{G}_{0}^{-1\dagger} 
], \label{eq:torquetracewithC}
\end{align}
where $\hat{\mathbf{J}} = \hat{\mathbf{L}} + \hat{\mathbf{S}}$ is the
total angular momentum operator and the trace is taken over the
vector components and position arguments. The final line follows by taking an ensemble average $\langle\bm{\tau}\rangle$ of the torque, expanding the integrand, and replacing $\langle E_{a}(\mathbf{r},
\omega)E_{b}^{*}(\mathbf{r}', \omega ) \rangle$ by the field-field correlator $\mathbb{C}_{ab}(\mathbf{r},
\mathbf{r}').$ The notation $|_{V_{body}}$ denotes
that the outer-most indices of the operator are traced over positions
in the body, while all others are over all space. The operator
$\mathbb{G}_{0}$ represents the background Green's function, which in
vacuum satisfies $\left[ \nabla\times\nabla\times -
  \frac{\omega^{2}}{c^{2}}\mathbb{I}\right]\mathbb{G}_{0}(\mathbf{r},
\mathbf{r}') = \mathbb{I}\delta^{(3)}(\mathbf{r} - \mathbf{r}').$ All
of the statistical properties of the sources in
Eq.~\eqref{eq:torquetracewithC} are represented by the field--field
correlation Dyadic $\mathbb{C}$ which, in the out of equilibrium
setting, can be decomposed $\mathbb{C}^{neq}(T_{env}, T_{body}) =
\mathbb{C}^{eq}(T_{env}) + [\mathbb{C}_{body}^{src}(T_{body}) -
  \mathbb{C}_{body}^{src}(T_{env})]$ as a sum of an equilibrium
$\mathbb{C}^{eq}(T_{env})$ plus a non-equilibrium term stemming from
the difference of the temperatures of the body and environment, with
the contribution due to the sources in the body (as opposed to the
environment) at a local temperature $T$ given
by~\cite{kruger_trace_formulae_for_nonequilibrium},
\begin{multline*}
    \mathbb{C}_{body}^{src}(T) =
\text{sgn}(\omega)\frac{\hbar\omega^{2}}{\pi
  c^{2}\epsilon_{0}} n(|\omega|,T)
\mathbb{G}_{0}\left(\mathbb{T}^{\mathsf{A}} -
\mathbb{T}\mathbb{G}_{0}^{\mathsf{A}}\mathbb{T}^{\dagger}
\right)\mathbb{G}_{0}^{\dagger},
\end{multline*}
where $n(\omega,T) = \frac{1}{\exp(\frac{\hbar\omega}{ k_{B}T}) - 1}$
is the Bose--Einstein distribution function. The scattering
$\mathbb{T}$ operator introduced above transforms incident fields into
induced currents in the body~\cite{rahi2009scattering}, and is
formally defined by the relation $\mathbb{T} = \mathbb{V}(\mathbb{I} -
\mathbb{G}_{0}\mathbb{V})^{-1}.$ Plugging the field--field
correlator $\mathbb{C}^{neq}(T_{env}, T_{body})$ into
Eq.~\eqref{eq:torquetracewithC} yields
\begin{align}
    \boldsymbol{\tau} &= -\textrm{Im}\int_{0}^{\infty} d\omega\,
    [n(\omega,T_{body}) - n(\omega,T_{env})] \nonumber \\
    &\qquad\qquad\times\frac{2}{\pi}\text{Tr}[\vecop{J}\mathbb{G}_{0}\big(\mathbb{T}^{\mathsf{A}} - \mathbb{T}\mathbb{G}_{0}^{\mathsf{A}}\mathbb{T}^{\dagger}\big)] \label{eq:pretorquetracesinglebody}\\
    &= 
    \int_{0}^{\infty} d\omega
    [n(\omega, T_{body}) - n(\omega, T_{env})]
    \nonumber\\
    &\qquad\qquad\times\underbrace{\frac{2}{\pi}
    \text{Tr}
    [(-\vecop{J}\mathbb{G}_{0}^{\mathsf{A}})\big(\mathbb{T}^{\mathsf{A}} - \mathbb{T}\mathbb{G}_{0}^{\mathsf{A}}\mathbb{T}^{\dagger}\big)
    ]}_{\Phi_{J}(\omega)}
    \label{eq:torquetracesinglebody}
\end{align}
The Tr symbol denotes a trace over the complete set of indices of the
enclosed operators (for example, both the position and polarization
indices of the dipole sources). The switch from
$\text{Tr}|_{V_{body}}$ to $\text{Tr}$ is possible since
$\mathbb{C}_{body}^{src}\mathbb{G}_{0}^{-1\dagger}$ has $\mathbb{T}$
or $\mathbb{T}^{\dagger}$ on the left or right of each term in the
expansion. As $\mathbb{T}$ vanishes for all points outside $V_{body}$,
one can extend the spatial integration to be over all space, resulting
in a trace expression. Furthermore, in going to the final expression
above, we used the Hermiticity of the quantity in parenthesis and
assumed a background environment with rotational symmetry (so that
$\mathbb{G}_{0}$ and $\vecop{J}$ commute) in which case, since $\vecop{J}$ is Hermitian,
$(\vecop{J}\mathbb{G}_{0})^{\mathsf{A}} =
\vecop{J}\mathbb{G}_{0}^{\mathsf{A}}.$ The assumption that
$\mathbb{G}_{0}$ describes a rotationally symmetric background is the
only symmetry assumption needed to arrive at the final expression
Eq.~\eqref{eq:torquetracesinglebody}. In particular, note that
$\mathbb{V}$ can be anisotropic or nonreciprocal.

The purely algebraic quantity $\Phi_{J}$ depends only on geometric and
material properties and can be directly interpreted as angular
momentum exchanged between the object and its environment, with
$-(\mathbb{T}^{\mathsf{A}} -\mathbb{T}
\mathbb{G}_{0}^{\mathsf{A}}\mathbb{T}^{\dagger})\hat{\mathbf{J}}$
describing absorption expressed as the subtraction of scattered
angular momentum from net extracted (extinction) angular momentum:
namely, the first term quantifies angular momentum extracted from an
incident wave upon interaction with the body while the second
describes angular momentum carried away by the scattered field. Note
that since fluctuations at different frequencies are uncorrelated, as
per Eq.~\eqref{eq:FDT}, the total rate of angular momentum transfer is
therefore given by an integral over all frequencies, with each
frequency contribution weighted by a difference of thermal occupation
numbers.

In general, the calculation of trace expressions for forces in the
basis of vector spherical harmonics (VSH) is complicated by the fact
that the matrix representation of $\hat{\mathbf{p}}$ is not diagonal
in this basis~\cite{khersonskii1988quantum}. Introducing multiple
bodies adds further complications. However, beyond its logical
necessity, the appearance of the total angular momentum $\hat{J}_{z}
\equiv \hat{L}_{z} + \hat{S}_{z}$ in torque trace expressions offers
computational advantages for torque calculations compared to force
calculations. In the basis of VSH, $\hat{J}_{z}$ is
diagonal~\cite{khersonskii1988quantum}, suggesting that torque
calculations in certain physical setups might be simpler analytically
and numerically than the corresponding force calculations. As an
illustrative example, we first consider bounds on the maximum
non-equilibrium torque that a single compact body may experience.

\begin{figure*}[t]
 \includegraphics[width=1.0\textwidth]{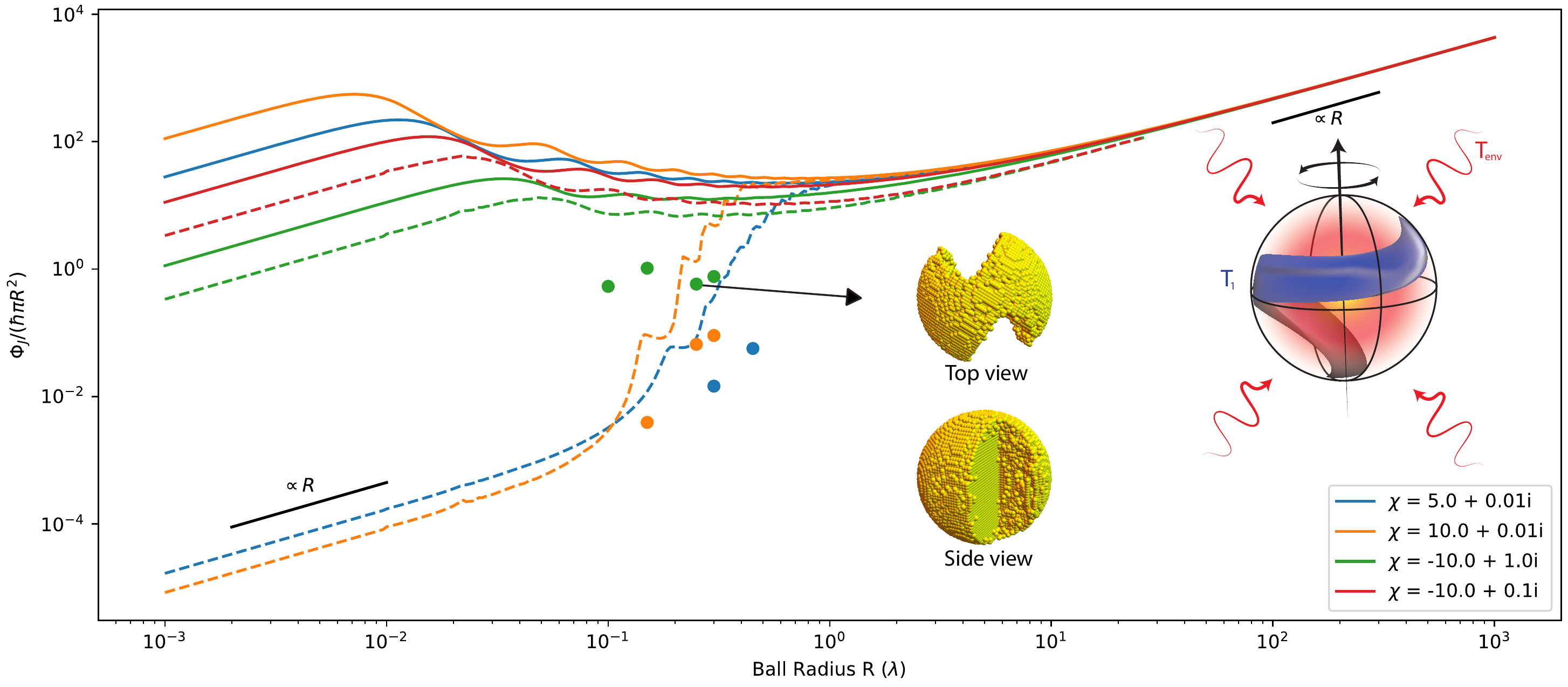} 
 \caption{\textbf{Bounds on maximum angular momentum transfer based on
     the conservation of energy.} An isolated body out of equilibrium can exchange net angular momentum with the environment, illustrated by an inset schematic. 
     The figure shows the maximum spectral angular momentum transfer $\Phi_J(\omega)$ (defined in the main text) at a single wavelength $\lambda$ allowed for a body enclosed in a spherical design volume of radius $R$, for multiple values of body material susceptibilities $\chi(\omega)$, captured by the material factor
     $\lVert\chi(\omega)\rVert^{2}/\Im\left[\chi(\omega)\right].$  $\Phi_{J,max}(\omega)$ is
   seen to smoothly blend the intuitively expected $\propto V$
   behaviors of quasi-static and ray-optic regimes, with an
   intermediate regime of growing radiative losses that suppresses the
   overall prefactor in the volumetric scaling.
   The full bounds, involving a relaxation of physics through the conservation of energy or optical theorem~\cite{jackson1999classical}, described in App.~\ref{app:torquebounds}, are shown as dashed lines. Solid lines describe a semi-analytical bound that, while looser owing to additional relaxations, provides an intuitive picture of the various wave contributions to torque (see main text).
   The dots indicate values of $\Phi_{J}(\omega)$ observed in topology optimized structures~\cite{polimeridis2014computation,polimeridis2015fluctuating,polimeridisgithub}.
   }
 \label{fig:torque_bounds}
\end{figure*}

\subsection{Size scaling and maximum torque}

Equation \eqref{eq:torquetracesinglebody} is valid for arbitrarily
structured objects of any size and shape. 
As the VSHs are eigenstates of $\hat{J}_{z},$ it is convenient to express the underlying scattering operators in a
spectral basis of VSHs~\cite{khersonskii1988quantum,tsang2000book}; choosing the
origin to lie at the center of mass, the vacuum Green's function can
be written as $\mathbb{G}_{0}^{\mathsf{A}} = \sum_{P,j,m}
\left|P,j,m\right> \left<P,j,m\right|$ (the eigenvalues are not
explicitly factored out, allowing for easier extraction of scaling
behavior below) and the net angular momentum exchange as
$\Phi_{J}(\omega) = -\frac{2\hbar}{\pi} \sum_{\substack{P j m \\ P' j'
    m'}}m ( \textrm{Im}[\mathcal{T}_{P j m; P' j' m'}] \delta_{P' j'
  m'; P j m} - \mathcal{T}_{P j m; P' j' m'}\mathcal{T}_{P j m; P' j'
  m'}^{*})$ in terms of the matrix elements of the $\mathbb{T}$
operator, $\mathcal{T}_{P j m; P' j' m'} \equiv \left<P, j, m\left|
\mathbb{T} \right|P', j', m'\right>$. It follows that objects for
which the scattering operator satisfies $\mathcal{T}_{P, j, m; P', j',
  m'} = \mathcal{T}_{P, j, -m; P', j', -m'}$ (for example, spherically
or cylindrically symmetrical bodies), exhibit zero Casimir torque,
owing to the lack of a preferred direction of radiation. Intuitively,
for small objects (particles), the scaling of the lowest order
scattering elements $\mathcal{T}_{N1m,N1m} \propto R^{3}$, with all
other matrix elements being higher order in $R$, yields a torque which
scales like the volume of the object. For larger sizes, the situation
becomes complicated owing to higher order scattering and the stronger
dependence on geometry. Thankfully, a recently developed formulation
of electromagnetic
bounds~\cite{global_T_operator_bounds,chao2022physical} allows
shape-agnostic analysis of size scaling which, perhaps not
surprisingly, reveals persistent volumetric scaling beyond quasitatic
settings~\footnote{Although the force $F$ and the torque $\tau$ may have different size scalings, we note that the linear acceleration $a = F/m$ and
  angular acceleration $\alpha = \tau/I,$ where $m$ is the mass $I$ is
  the moment of inertia, also having different size scalings in the
  denominators. Let $R$ denote the system size. The mass $m$
  scales like $R^{3}$, while $I \propto mR^{2}\propto R^{5}$.}.

Generally, the question of what kind of geometry leads to maximum
torque is interesting and can, in absence of intuitive
characteristics, be probed via large scale
optimization~\cite{molesky2018inverse,christiansen2021inverse}.  Further understanding
e.g. scaling behavior, can be achieved by applying a recent framework
based on Lagrange duality to compute shape-independent bounds,
previously used in the context of thermal
radiation~\cite{T_operator_bounds_angle_integrated,global_T_operator_bounds,chao2022physical}.
Concisely, and at a high level, bounds are obtained by maximizing a
desired objective function: the contribution to the torque
Eq.~\eqref{eq:torquetracesinglebody} at a single characteristic
angular frequency of the absorption spectrum of the object, with
respect to possible scattering operator response~\footnote{We are
  primarily interested in arbitrary designs within the prescribed
  region whose center of mass is at the origin. This may be viewed as
  a relaxation of a `center of mass' constraint.} subject to
constraints incorporating a subset of the scattering physics of the
problem. For simplicity, we consider non-magnetic materials ($\mu =
\mathbb{I}$). Supposing a local isotropic material susceptibility ($\chi$)
and isotropic background environment ($\mathbb{G}_{0}$),
Eq.~\eqref{eq:torquetracesinglebody} becomes rotationally
invariant. Accordingly, both the chosen direction and sign of the
objective $\Phi_{J}$---the geometry dependent component of
Eq.~\eqref{eq:torquetracesinglebody}---are immaterial to the
optimization; the optimal values for the maximization and minimization
of $\Phi_{J}$ differ by a minus sign.

Maximizing $\Phi_J$ by considering the optimal $\mathbb{T}$ can be
achieved by moving to an eigenbasis of $\mathbb{G}_{0}^{\mathsf{A}}.$
In particular, since $\mathbb{G}_{0}^{\mathsf{A}}$ describes radiation
away from an object into the surrounding
environment~\cite{landau2013statistical, molesky2019bounds}, an
eigenbasis of $\mathbb{G}_{0}^{\mathsf{A}}$ is a natural choice to
evaluate the trace. Furthermore, the vector spherical harmonics are,
by definition, eigenstates of $\hat{J}_{z}$ (see the appendix for a
review). Working in this eigenbasis of
$\mathbb{G}_{0}^{\mathsf{A}}$~\cite{tsang2000book}, one finds
$\hat{J}_{z}\mathbb{G}_{0}^{\mathsf{A}}(\mathbf{x}, \mathbf{y}) =
k\sum_{j,m}(-1)^{m} m\hbar [
  \mathbf{RM}_{j,m}(k\mathbf{x})\mathbf{RM}_{j,-m}(k\mathbf{y}) +
  \mathbf{RN}_{j,m}(k\mathbf{x})\mathbf{RN}_{j,-m}(k\mathbf{y})].  $
This choice of basis is further useful and natural if the design
domain is a spherical ball (the body must fit within a sphere
of radius $R$), as will be the case in this article.  In order to keep
the expressions more compact, we will write the eigenmode
expansion~\footnote{The extensive use of the eigenvalues of
  $\mathbb{G}_{0}^{\mathsf{A}}$ in optimization analysis (writing code
  and analytical work) proved it convenient to use basis elements
  normalized over the design domain e.g. a sphere of radius $R.$ Other
  conventions absorb the $\rho_{n}$ into the definition of the basis
  vectors.}  as $ \mathbb{G}_{0}^{\mathsf{A}} = \sum_{n}
\rho_{n}\left|\mathbf{Q}_{n}\right> \left<\mathbf{Q}_{n}\right|,
\label{asymGexpansion}
$ where each radiative coefficient $\rho_{n}$ is non-negative due to
passivity (that is, $\mathbb{G}_{0}^{\mathsf{A}}$ is positive
semi-definite). Therefore, the eigenvectors $\ket{\mathbf{Q}_{n}}$ can
be indexed as $\ket{\mathbf{Q}_{P, j, m}},$ where $P$ denotes the type
(M or N wave), $j = 1, 2, \dots$, and $m = -j, -j + 1, \dots, j-1, j,$
and $\hat{J}_{z}|\mathbf{Q}_{P, j, m}\rangle = m\hbar |\mathbf{Q}_{P,
  j,m}\rangle.$ The eigenvalues can be expressed in the case of a
sphere of radius $R$ (for orthonormal basis vectors) as
\begin{align}
    \rho_{M,j,m} &= \frac{\pi (kR)^{2}}{4k^{2}}\left(J_{j + \frac{1}{2}}^{2}(kR) - J_{j - \frac{1}{2}}(kR)J_{j + \frac{3}{2}}(kR)  \right)
\end{align}

\begin{align}
    \rho_{N,j,m} &= \frac{\pi (kR)^{2}}{4k^{2}} \times \nonumber \\ &\hspace{-0.3in}\Bigg[
    \frac{j+1}{2j+1}\bigg(J_{j - \frac{1}{2}}^{2}(kR) - J_{j + \frac{1}{2}}(kR)J_{j 
    - \frac{3}{2}}(kR)\bigg) \nonumber \\
    &\hspace{-0.2in}+\frac{j}{2j+1}\left(J_{j + \frac{3}{2}}^{2}(kR) - J_{j + \frac{1}{2}}(kR)J_{j + \frac{5}{2}}(kR)\right)
    \Bigg]
\end{align}
where $J_{\nu}$ is a Bessel function of the first kind of order $\nu.$

In this basis and setting $\left|\mathbf{T}_{n}\right> \equiv
\mathbb{T}\left |\mathbf{Q}_{n}\right>$, with
$-\frac{i}{kZ}\left|\mathbf{T}_{n}\right>$ denoting the electric
polarization current density in the object resulting from the $n$-th
radiative mode, one finds
 \begin{equation}
     \Phi_{J} = -\frac{2}{\pi}\sum_{n}\rho_{n}\left(\Im\left[
     \left<\mathbf{Q}_{n}|\hat{J}_{z}|\mathbf{T}_{n}\right>\right] - 
     \left<\mathbf{T}_{n}\right|\hat{J}_{z}\mathbb{G}_{0}^{\mathsf{A}}
     \left|\mathbf{T}_{n}\right>\right).
     \label{phiSum}
 \end{equation}
The form of $\Phi_{J}$ implies that there is a limit on the
torque. The argument is similar to that of bounds for power
quantities~\cite{miller2016fundamental} and relies on the competition
between the linear and quadratic terms in the polarization currents
which limits the magnitude of the optimal polarization current.  In
particular, note also that in this basis, $\hat{J}_{z}$ and
$\hat{J}_{z}\mathbb{G}_{0}^{\mathsf{A}}$ each break down into a
positive definite block ($m > 0$), a 0 block ($m = 0$), and a
negative-definite block ($m < 0$). Noting the overall minus sign in
$\Phi_{J}$, therefore, in order to optimize absorption it is clear
from Eq.~\eqref{phiSum} that each radiative mode within the
negative-definite block must generate a strong polarization:
$-\Im\left[
  \left<\mathbf{Q}_{n}|\hat{J}_{z}|\mathbf{T}_{n}\right>\right]$ is
the extracted angular momentum.  However, the generation of these
currents necessarily leads to radiative losses of angular momentum,
$-\left<\mathbf{T}_{n}\right|\hat{J}_{z}
\mathbb{G}_{0}^{\mathsf{A}}\left|\mathbf{T}_{n}\right>$, which grow
relatively in strength as the size of the domain increases through the
growth of the $\rho_{n}$ radiative coupling
coefficients~\cite{molesky2019bounds,venkataram2020fundamental}. Likewise,
each radiative mode within the positive-definite block should ideally
not generate a polarization current.

At the coarsest level of this relaxation procedure, with details in
the appendix, the loosest $\Phi_{J,max}$, consistent only with optimal
scattering satisfying not the full scattering equations but merely the
conservation of power (optical theorem~\cite{jackson1999classical})
over the entire domain, is
\begin{align}
 \Phi_{J, max}
 = \frac{\hbar}{2\pi}\sum_{P,j,m}
 \begin{cases}
   ~0 & \text{} m \geq 0 \\
   \begin{cases}
            -m &\text{} \zeta\rho_{P,j,m} > 1 \\
            -\frac{4m\zeta\rho_{P,j,m}}{(1 + \zeta\rho_{P,j,m})^{2}} &\text{} \zeta\rho_{P,j,m} \leq 1 \\
    \end{cases}
    & \text{} m < 0 \\
 \end{cases}
 \label{realPowerSol}
\end{align}
where $\zeta \equiv k^{2}\lVert\chi\rVert^{2}/\Im\left[\chi\right]$ is
a measure of the dissipative response of the
system~\cite{miller2016fundamental,molesky2019bounds}. The simplicity
of $\Phi_{J,max}$ as arising from a sum over independent channel
contributions lends itself to simple interpretation.  The optimal
polarization currents associated with maximum angular momentum
transfer in each channel can be chosen to be proportional to the
radiative states, here taken to be the eigenbasis of
$\mathbb{G}_{0}^{\mathsf{A}}$, such that
$\mathbb{T}\ket{\vec{Q}_{P,j,m}} =
c_{P,j,m}\left|\mathbf{Q}_{P,j,m}\right>$, with the maximum bound
polarization response for each channel set by $\lVert c_{P,j,m}\rVert
\leq \min\left\{\frac{1}{2\rho_{P,j,m}}, \zeta\right\}$ with $m <0$
and $c_{P,j,m} = 0$ for $m \geq 0.$

Figure~\ref{fig:torque_bounds} shows $\Phi_{J, max}$ for various
system parameters, illustrating the dependence of maximal
angle-integrated angular momentum transfer for bodies of different
shapes and material compositions enclosed in a spherical ball of radius $R.$
It is observed that the mere imposition of energy
conservation is sufficient for the bounds to show intuitive
quasi-static and ray-optic behavior. In the limit of a small design
volume, $\zeta\rho_{P,j,m} \ll 1$ for all $\{P,j,m\}$, $\Phi_{J, max}$
is seen to exhibit volumetric scaling consistent with the assumption
that the magnitude of all generated polarization currents can grow as
large as material loss allows: as the volume grows, so does the
available angular momentum in each channel, and hence so should the
polarization response. Intuitively, if the object size $R$ is smaller
than the penetration (skin) depth of the medium, then one expects the
entire volume of the object to interact with any impinging waves.
Owing to the necessary coupling of the currents with radiative waves,
as $R$ increases there is a decrease in how much net angular momentum
can be transferred per volume of the object.  In the intermediate
regime where the object is on the wavelength scale, in each index of
Eq.~\eqref{phiSum} growth in $\rho_{P,j,m}$ causes radiative losses to
compete with the net extracted angular momentum if the magnitude of
$\left|\mathbf{T}_{P,j,m}\right>$ becomes too large, leading the
associated channel (index) to enter the saturation condition of
Eq.~\eqref{realPowerSol}, visible in Fig.~\ref{fig:torque_bounds} as
the onset of steps. As an increasing number of channels saturate, the
volumetric scaling appears to transition to area scaling as observed
for radiated power~\cite{molesky2019bounds}, but only temporarily.
Intuitively, if the object size $R$ is significantly larger than the
penetration depth, the effective portion of the object interacting
with an impinging wave is expected to scale like the surface area
times the penetration depth (which is material dependent, but
independent of the object size). The angular momentum for the photons
on the surface relative to the center of mass of the object is
expected to have orbital contributions which scale like the distance
from the origin, suggesting $R \times R^{2}\propto V$ scaling
again. Consequently, one finds that spin and orbital contributions
each dominate in the quasistatic and ray-optic limits, respectively,
leading to volumetric scaling in either regime.

As support for the above intuitive picture, the small $R$ asymptotic
(point particle limit) for $\Phi_{J}$ can be carried out analytically
(see appendix) to yield,
\begin{align}
  &\boldsymbol{\tau}(T_{body}, T_{env}) \nonumber \\
    &= 
    -\frac{2}{\pi}
    \int_{0}^{\infty} d\omega
    [n(\omega, T_{body}) - n(\omega, T_{env})]
    \nonumber\\
    &\times\text{Tr}
        [
        \frac{\omega^{3}}{6\pi c^{3}}\vecop{S}\bigg(
        \MatrixVariable{\alpha}^{\mathsf{A}}
        -
        \frac{\omega^{3}}{6\pi c^{3}}\MatrixVariable{\alpha} \hspace{1px}\MatrixVariable{\alpha}^{\dagger}
        \bigg)
        ]
    \label{eq:torquetracesinglePP},
\end{align}
where $\MatrixVariable{\alpha}(\omega) \equiv 4\pi
R^{3}\frac{\MatrixVariable{\epsilon}(\omega) -
  \mathbb{I}}{\MatrixVariable{\epsilon}(\omega) + 2\mathbb{I}}$ is the
polarizability of the particle and Tr is a trace of a 3-by-3 matrix.
Notably, the $R^{3}$ order terms from the $\hat{L}_{z}$ operator
vanish exactly leaving only the $\hat{S}_{z}$ dependence. (Note
furthermore that for a reciprocal particle and to linear order in
volume, the torque vanishes exactly.)

\section{Casimir Torque for Multiple Bodies}

Although we have so far focused on the case of a single body, the
trace expressions derived above can be extended to incorporate
interactions between multiple objects at different temperatures. The
analysis follows a similar approach to that of nonequilibrium heat
transfer and force described in
Refs.~\cite{kruger_trace_formulae_for_nonequilibrium,bimonte2016non},
so below we simply summarize the salient points. Suppose that
there is a set of $N$ bodies (not counting the environment) indexed by
$\alpha = 1, 2, \dots, N.$ The starting point is
Eq.~\eqref{eq:torquetracewithC}, where the volume integral is over
object $\alpha.$ The total Casimir torque on the $\alpha^{th}$ object
can be decomposed as
\begin{align}
    \boldsymbol{\tau}^{(\alpha),neq}(T_{env}, \{T_{\beta}\}) &= \boldsymbol{\tau}^{(\alpha), eq}(T_{env}) + \nonumber \\ &\sum_{\beta}[\boldsymbol{\tau}_{\beta}^{(\alpha)}(T_{\beta}) - \boldsymbol{\tau}_{\beta}^{(\alpha)}(T_{env})].
    \label{eq:torqueeqplusneq}
\end{align}
That is, the total Casimir torque in nonequilibrium can be written as
a sum of an equilibrium contribution $\boldsymbol{\tau}^{(\alpha),
  eq}(T_{env})$ where all objects are at a temperature $T_{env}$ plus
non-equilibrium contributions when the objects $1, \dots, N$ deviate
from the temperature of the background environment $T_{env}.$ This
follows from the field--field correlator in non-equilibrium $
\mathbb{C}^{neq}(T_{env}, \{T_{\beta}\}) = \mathbb{C}^{eq}(T_{env}) +
\sum_{\beta} [\mathbb{C}_{\beta}^{src}(T_{\beta}) -
  \mathbb{C}_{\beta}^{src}(T_{env})]$ which has an equilibrium
correlation part and a sum of terms that measure the contributions to
the field--field correlator for objects held at different temperatures
from the background
environment~\cite{kruger_trace_formulae_for_nonequilibrium}. $\boldsymbol{\tau}_{\beta}^{(\alpha)}(T)$
is the torque on $\alpha$ due to sources in $\beta,$ when body $\beta$
is at a temperature $T$ and $\mathbb{C}^{src}_{\beta}(T)$ denotes the
contribution to the field--field Dyadic from sources in object $\beta$
and scattered by all other objects.  Calculating the torque on object
$\alpha$ due to $\mathbb{C}_{\beta}^{src}(T)$ involves a spatial
integral $\int_{V_{\alpha}} d^{3}\mathbf{r}(\dots)$ only over the
volume of $\alpha,$ which is not a trace expression. However, further
analysis carried out in
Ref.~\cite{kruger_trace_formulae_for_nonequilibrium} in the case of
force calculations and omitted here proves that one can indeed extend
the integral to the entire domain, resulting in a basis-independent trace expression.
Carrying out a similar procedure in the case of two bodies yields
\begin{widetext}
    \begin{align}
    \boldsymbol{\tau}_{1}^{(1)}(T) &= -\frac{2}{\pi} \int_{0}^{\infty} d\omega ~  n(\omega, T)
    \text{Im}\text{Tr}[
    \vecop{J}(1 + \mathbb{G}_{0}\mathbb{T}_{2})
    \frac{1}{1-\mathbb{G}_{0}\mathbb{T}_{1}\mathbb{G}_{0}\mathbb{T}_{2}}
    \mathbb{G}_{0}
    (\mathbb{T}_{1}^{\mathsf{A}} - \mathbb{T}_{1}\mathbb{G}_{0}^{\mathsf{A}}\mathbb{T}_{1}^{\dagger})
    \frac{1}{1-\mathbb{G}_{0}^{\dagger}\mathbb{T}_{2}^{\dagger}\mathbb{G}_{0}^{\dagger}\mathbb{T}_{1}^{\dagger}}
    ], \label{eq:tau11}\\
    \boldsymbol{\tau}_{2}^{(1)}(T) &= -\frac{2}{\pi} \int_{0}^{\infty} d\omega ~  n(\omega, T)
    \text{Im}\text{Tr}[
    \vecop{J}(1 + \mathbb{G}_{0}\mathbb{T}_{1})
    \frac{1}{1-\mathbb{G}_{0}\mathbb{T}_{2}\mathbb{G}_{0}\mathbb{T}_{1}}
    \mathbb{G}_{0}
    (\mathbb{T}_{2}^{\mathsf{A}} - \mathbb{T}_{2}\mathbb{G}_{0}^{\mathsf{A}}\mathbb{T}_{2}^{\dagger})\mathbb{G}_{0}^{\dagger}
    \frac{1}{1-\mathbb{T}_{1}^{\dagger}\mathbb{G}_{0}^{\dagger}\mathbb{T}_{2}^{\dagger}\mathbb{G}_{0}^{\dagger}}
    \mathbb{T}_{1}^{\dagger}
    ].
    \label{eq:tau21}
\end{align}
\end{widetext}
Plugging these expressions into Eq.~\eqref{eq:torqueeqplusneq}, one
can thus obtain the various contributions to the torque on either
object. Note that the above equations follow directly from
Eqs.~(76--77) in Ref.~\cite{kruger_trace_formulae_for_nonequilibrium}
upon the substitution $\hat{\mathbf{p}} \to \hat{\mathbf{J}}$,
corresponding to the change in observable from linear momentum
$\hat{\mathbf{p}} = -i\hbar\nabla$ to \emph{net} angular momentum as
derived and discussed in Sec.~I.



As in the section above and for illustrative purposes, we now consider
the special case of two point particles (radius smaller than any other
length scale in the problem including the thermal wavelength, skin
depth, and inter-particle distance) of polarizabilities
$\MatrixVariable{\alpha}_{1}$ and $\MatrixVariable{\alpha}_{2}$ held
at temperatures $T_{1}$ and $T_{2}$ compared to a vacuum environment
of temperature $T_{env}$.  In this limit, one can neglect multiple
scatterings and work within the Born approximation so that only the
lowest order terms in the scattering operators are kept, in which case
the inverse operators $\frac{1}{\mathbb{I} - \dots}$ in the trace
expression above become the identity and the above expressions
simplify to yield,
\begin{align}
        \boldsymbol{\tau}_{1}^{(1)}(T)\cdot\mathbf{e}_{z} &= -\frac{2}{\pi}
    \int_{0}^{\infty}d\omega \, \frac{\omega^{2}}{c^{2}} n(\omega,T) \times \nonumber \\ 
        &\hspace{-0.5in} \Bigg(\text{Tr}
        \Big[
        (\hat{J}_{z}\mathbb{G}_{0}^{\mathsf{A}})(\mathbf{r}_{1},\mathbf{r}_{1})\big(
        \MatrixVariable{\alpha}_{1}^{\mathsf{A}}
        -
        \frac{\omega^{2}}{c^{2}}\MatrixVariable{\alpha}_{1}\mathbb{G}_{0}^{\mathsf{A}}(\mathbf{r}_{1},\mathbf{r}_{1})\MatrixVariable{\alpha}_{1}^{\dagger}
        \big)
        \Big] \nonumber \\
        &\hspace{-0.4in} + \frac{\omega^{2}}{c^{2}} \text{Im}\text{Tr}
        \Big[
        (\hat{J}_{z}\mathbb{G}_{0})(\mathbf{r}_{1},\mathbf{r}_{2})
        \MatrixVariable{\alpha}_{2}\mathbb{G}_{0}(\mathbf{r}_{2},\mathbf{r}_{1})\MatrixVariable{\alpha}_{1}^{\mathsf{A}}
        \Big]\Bigg) \\
    \boldsymbol{\tau}_{2}^{(1)}(T)\cdot\mathbf{e}_{z} &= -\frac{2}{\pi}
    \int_{0}^{\infty}d\omega \, \frac{\omega^{4}}{c^{4}} n(\omega,T) \times \nonumber\\
    &\text{Im}\text{Tr} \Big[(\hat{J}_{z}\mathbb{G}_{0})(\mathbf{r}_{1},
      \mathbf{r}_{2})\MatrixVariable{\alpha}^{\mathsf{A}}_{2}\mathbb{G}_{0}^{\dagger}(\mathbf{r}_{2},
      \mathbf{r}_{1})\MatrixVariable{\alpha}_{1}^{\dagger}\Big]
    \label{eq:tauPP2onPP1simplified}
\end{align}
where $\mathbf{r}_{1}, \mathbf{r}_{2}$ are the locations of the point
particles and the trace involves only a sum over the vector components
so that $\text{Tr}[\mathbb{A}(\mathbf{r}, \mathbf{r}')] \equiv
\sum_{a} \mathbb{A}_{aa}(\mathbf{r}, \mathbf{r}')$. 

\begin{figure}
  \includegraphics[width=0.48\textwidth]{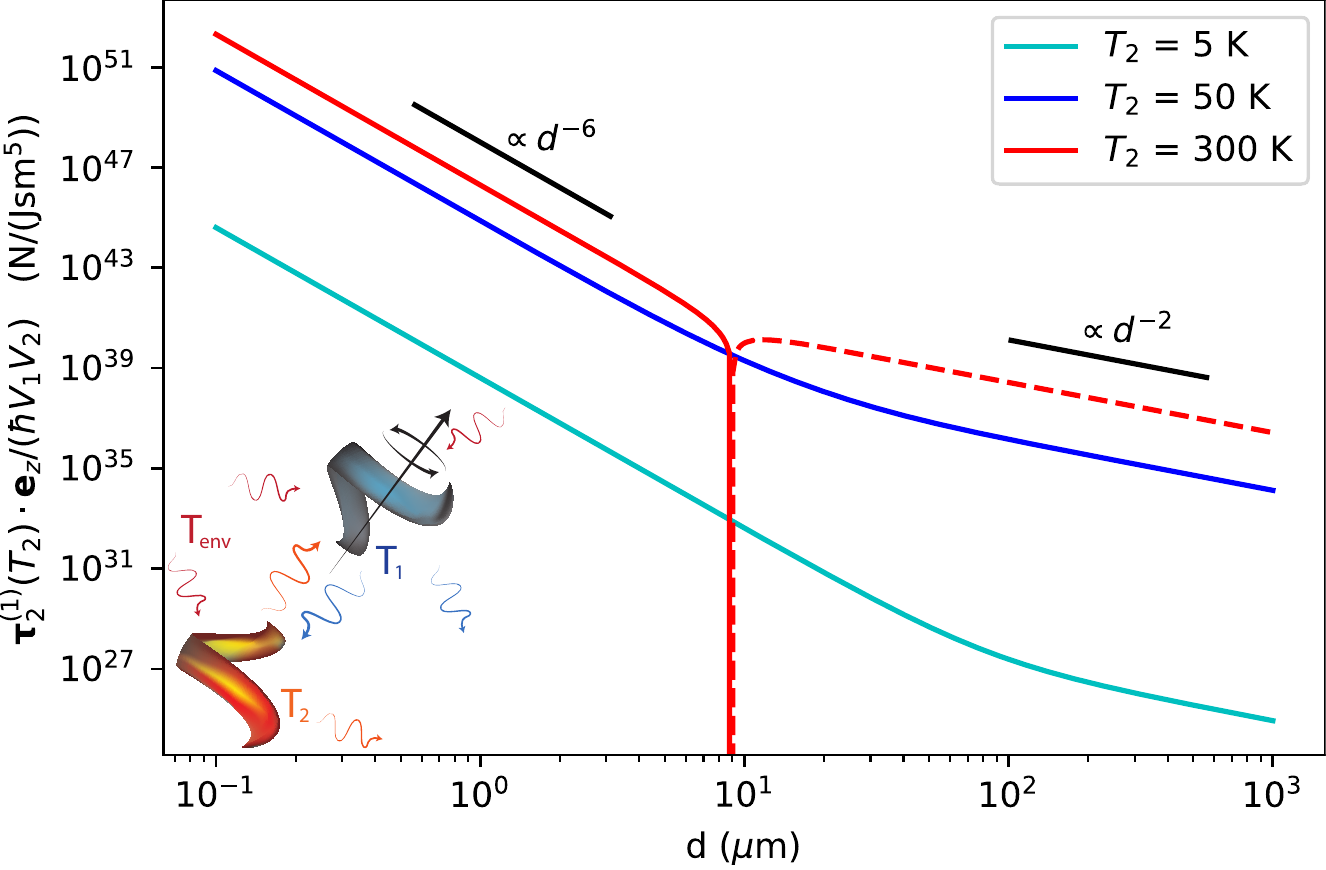}
     \caption{\textbf{Torque on a particle due to fluctuating sources in a neighboring particle.} Two objects can exchange net angular momentum amongst themselves and their environment, illustrated by an inset schematic. For concreteness, we plot $\boldsymbol{\tau}_{2}^{(1)}(T_{2})\cdot\mathbf{e}_{z}$, the torque on particle 1 due to the sources in particle 2,
     normalized by $\hbar V_1 V_2$ where $V_{1}$ and $V_{2}$ are the volumes of particle 1 and 2, respectively, as
     a function of the separation $d$ along the $x$-direction of two InSb particles in the $xy$-plane subject
     to an external magnetic field in the $z$-direction. Solid/dashed lines indicate positive/negative
     values.}
     \label{fig:torqueTwoPPs}
\end{figure}

Let $V_{1}$ and $V_{2}$ denote the volumes of particles 1 and 2, respectively, which are separated by a distance $d.$ Since the vacuum Green's functions scale as $1/d^{3}$ and $1/d$ in the
near- and far-fields, respectively, one expects the separation-dependent parts of both quantities to
scale $\propto V_{1}V_{2}/d^{6}$ and $\propto V_{1}V_{2}/d^{2}$ for
small and large separations, respectively. Note however the presence
of a separation-independent term in $\boldsymbol{\tau}_{1}^{(1)}\cdot\mathbf{e}_{z}
\propto V_{1}$. As in the single-body case, one can show (see
appendix) that $(\hat{J}_{z}\mathbb{G}_{0}^{\mathsf{A}})(\mathbf{r},
\mathbf{r}) = (\hat{S}_{z}\mathbb{G}_{0}^{\mathsf{A}})(\mathbf{r},
\mathbf{r}) = \frac{k}{6\pi}\hat{S}_{z}$, leading to vanishing torque
on reciprocal particles, $\MatrixVariable{\alpha}_{1,xy} =
\MatrixVariable{\alpha}_{1,yx}$, to leading order in their
volumes. Plugging the various torque contributions into
Eq.~\eqref{eq:torqueeqplusneq}, one finds that as $d\to\infty,$ the
net torque $\boldsymbol{\tau}^{(1),neq}$ on particle 1 is dominated by the
separation-independent self-torque $\boldsymbol{\tau}_{1}^{(1)}$ derived
in the previous section and given by
Eq.~\eqref{eq:torquetracesinglePP}. For a concrete example
illustrating the above salient features we consider the torque arising
in a configuration of two InSb particles subjected to an external magnetic field of magnitude $1$ T ($10^{4}$
Gauss), resulting in a permittivity of the form
\begin{align}
    \MatrixVariable{\epsilon} =
    \begin{bmatrix}
    \epsilon_{1} & -i\epsilon_{2} & 0 \\
    i\epsilon_{2} & \epsilon_{1} & 0 \\
    0 & 0 & \epsilon_{3}
    \end{bmatrix}_{cart}.
\end{align}
Figure~\ref{fig:torqueTwoPPs} shows
$\boldsymbol{\tau}_{2}^{(1)}(T_{2})\cdot\mathbf{e}_{z}$, normalized by
$\hbar V_1 V_2$, as a function of separation of two particles of
radius 100 nm in a zero-temperature vacuum and held at
different temperatures $T_1 = 0$ K and $T_2$. In addition to showcasing
the above-mentioned scaling with $d$, the plots illustrate that the temperatures and separations determine which frequency contributions to the total torque dominate, resulting in possible
transitions in the \emph{sign} of the torque~\cite{kruger2011spheresandplates}.
For the settings shown, the largest possible torque is roughly $10^{-23}$ Nm, occurring for $T_{2} = 300$ K and $d \approx
100$~nm.
Dividing by the moment of inertia of the particle ($\frac{2}{5}M_{1}R_{1}^{2}$) yields a
potential angular acceleration for particle 1 around its center of
mass of roughly $400$ rad/s$^{2}$ at $d \approx 1 \ \mu$m.

\section{Concluding Remarks}

In summary, we have introduced trace expressions for non-equilibrium
Casimir torque that apply to arbitrary object shapes and materials,
generalizing prior work on power transfer~\cite{T_operator_bounds_angle_integrated,kruger_nonequilbrium_fluctuations_review,venkataram2020fundamental} and forces~\cite{kruger_nonequilbrium_fluctuations_review,kruger_trace_formulae_for_nonequilibrium,venkataram2020fundamentalCPforce} and showing
explicitly the need for a full account of the spin and orbital angular
momentum carried by waves in this setting. Furthermore, we have shown
that recently developed
techniques~\cite{molesky2020t,molesky2020hierarchical,molesky2021comm,chao2022physical,venkataram2020fundamentalCPforce,venkataram2020fundamental}
for calculating bounds on sesquilinear objectives in electromagnetics
can be applied to torque problems,
revealing volumetric scaling for small and large object asymptotics in
a shape-independent framework. The closeness of the associated limits
with specific body shapes as discovered by inverse design, continues a
trend observed in previous works on bounds to thermal absorption and
emission~\cite{global_T_operator_bounds,molesky2020hierarchical}. Although the calculated
bounds focused exclusively on contributions from a dominant frequency,
extensions to net (spectrally integrated) torque can
be carried out as
described in Ref.~\cite{molesky2021comm}. Further extensions to analyze the impact of
nonreciprocal and anisotropic media will also be considered in the
near future.

\begin{acknowledgments}

This work was supported by the National Science Foundation under the Emerging Frontiers in Research and Innovation (EFRI) program, EFMA-1640986, the Cornell Center for Materials Research (MRSEC) through award DMR-1719875, the Defense Advanced Research Projects Agency (DARPA) under agreements HR00112090011, HR00111820046 and HR0011047197, and the Canada First Research Excellence Fund via the Institut de Valorisation des Donn\'{e}es (IVADO) collaboration. The views, opinions, and findings expressed herein are those of the 
authors and should not be interpreted as representing the official 
views or policies of any institution. 
\end{acknowledgments}

\appendix

\section{Thermal field correlations in equilibrium and non-equilibrium settings}

In this section, we summarize the details of
the use of fluctuation-dissipation theorem (FDT) and local equilibrium properties with the scattering
equations as described
in Refs.~\cite{kruger_nonequilbrium_fluctuations_review,kruger_trace_formulae_for_nonequilibrium}, modified to SI units and without assumptions of reciprocity.
Let $\langle\cdots\rangle_{T}$ denote an equilibrium thermal average
at a temperature $T.$ In the Rytov formalism for fluctuation
electrodynamics~\cite{rytov1989principles}, one assumes that the free dipoles inside an object
fluctuate. In thermal equilibrium, one can show that the
fluctuation-dissipation theorem
gives~\cite{novotny2012principles,kruger_nonequilbrium_fluctuations_review},
  \begin{align}
 &\underbrace{\left\langle J_{i}(\mathbf{x}, \omega)
 J_{j}^{*}(\mathbf{x}', \omega ') 
 \right\rangle_{T}}_{\text{fluctuation}} = \nonumber
 \\
 \label{fluctuationDissipation}
 &\frac{\omega\epsilon_{0}}{2\pi} 
 \coth\left( \frac{\hbar
 \omega}{2k_{B}T} \right)
 \underbrace{\bm{\chi}_{ij}^{\mathsf{A}}(\mathbf{x},
 \mathbf{x}';\omega)
 }_{\text{dissipation}}
 \delta(\omega - \omega'),
\end{align}

\begin{align}
    \mathbb{C}_{ij}^{eq}(T, \omega,\omega'; \mathbf{r}, \mathbf{r}') &\equiv \left\langle E_{i}(\mathbf{r}, \omega)E_{j}^{*}(\mathbf{r}', \omega ') \right\rangle_{T} \\
    &= b(T)\delta(\omega - \omega') \mathbb{G}^{\mathsf{A}}_{ij}(\omega; \mathbf{r}, \mathbf{r}')
\end{align}
where
\begin{align}
    b(T) \equiv \frac{\hbar\omega^{2}}{2\pi c^{2}\epsilon_{0}} \coth\left( \frac{\hbar\omega}{2k_{B}T} \right).
\end{align}
One can also define
\begin{align}
    b(T) &\equiv a(T) + a_{zp} \\
    a(T) &\equiv \text{sgn}(\omega)\frac{\hbar\omega^{2}}{\pi 
    c^{2}\epsilon_{0}}\frac{1}{\exp\big(\frac{\hbar|\omega|}{k_{B}T} \big) - 1} \\ 
    a_{zp} &\equiv \text{sgn}(\omega)\frac{\hbar\omega^{2}}{2\pi c^{2}\epsilon_{0}}
\end{align}
to further break-up the terms into a temperature dependent piece and a quantum zero-point term. The Dyadic Green's function satisfies
\begin{align}
\label{eq:hemholtz}
    \left[  \mathbb{H}_{0} - \mathbb{V} - \frac{\omega^{2}}{c^{2}}\mathbb{I}\right]\mathbb{G}(\mathbf{r}, \mathbf{r}') = \mathbb{I}\delta^{(3)}(\mathbf{r} - \mathbf{r}'),
\end{align}
where \(\mathbb{H}_{0} = \nabla\times\nabla\times\) and 
\begin{align}
    \mathbb{V} = \frac{\omega^{2}}{c^{2}}(\mathbb{\epsilon} - \mathbb{I}) + \nabla\times(\mathbb{I} - \mathbb{\mu}^{-1})\nabla\times.
\end{align}
The vacuum Green's function $\mathbb{G}_{0}$ is the solution of Eq.~\eqref{eq:hemholtz} with $\mathbb{V} = 0.$

Before considering non-equilibrium situations, rewrite the equilibrium expressions. As a first step, use the (mathematically trivial) fact that \(\mathbb{G}^{\mathsf{A}} = -\mathbb{G}\mathbb{G}^{-1\mathsf{A}}\mathbb{G}^{\dagger}\) and the fact that, from Eq.~\eqref{eq:hemholtz},
\begin{align}
    -(\mathbb{G}^{-1} - \mathbb{G}_{0}^{-1})^{\mathsf{A}} = \mathbb{V}^{\mathsf{A}}.
\end{align}
This then lets one write
\begin{align}
    \mathbb{G}^{\mathsf{A}} = \mathbb{G}(\mathbb{V}^{\mathsf{A}} - \mathbb{G}_{0}^{-1\mathsf{A}})\mathbb{G}^{\dagger}
\end{align}

Suppose that there are \(N\) objects labeled by \(\alpha = 1,\dots, N.\) Then one may rewrite the expression for \(\mathbb{C}^{eq}\) to get
\begin{align}
    \mathbb{C}^{eq}(T) = \mathbb{C}^{zp} + \mathbb{C}^{env}(T) + \sum_{\alpha}\mathbb{C}_{\alpha}^{src}(T).
\end{align}
where \(\mathbb{V} = \sum_{\alpha} \mathbb{V}_{\alpha}\) and
\begin{align}
    \mathbb{C}_{zp} &= a_{zp} \mathbb{G}^{\mathsf{A}}, \\ 
    \mathbb{C}_{\alpha}^{src}(T) &= a(T)\mathbb{G} \mathbb{V}_{\alpha}^{\mathsf{A}}\mathbb{G}^{\dagger}, \\ 
    \mathbb{C}_{env}(T) &= -a(T)\mathbb{G}\mathbb{G}_{0}^{-1\mathsf{A}}\mathbb{G}^{\dagger}.
\end{align}
The interpretation is as follows. \(\mathbb{C}_{\alpha}^{src}\), which involves \(\mathbb{V}_{\alpha}\), is interpreted as the contribution to the spectral density function due to the sources in object \(\alpha\), so that the ``left-over" term \(\mathbb{C}^{env}\) is interpreted as the contribution from the environment (which is anything not described by the non-zero parts of \(\mathbb{V}\)).

At this point, some assumptions have to be made in order to proceed further. We assume that in the non-equilibrium situation one may still use the above decomposition by assuming that the fluctuations still satisfy the fluctuation-dissipation theorem at the corresponding local temperatures of each object. Also, we assume that the time scales for temperature changes are much longer than the time scales for observation of the mechanical effects. Assuming that all the temperatures are independently tunable, the non-equilibrium expression for the field correlator becomes
\begin{align}
\label{eq:Cneq}
    \mathbb{C}^{neq}(T_{env}, \{T_{\alpha}\}) &= \mathbb{C}_{zp} + \mathbb{C}_{env}(T_{env}) + \sum_{\alpha} \mathbb{C}_{\alpha}^{src}(T_{\alpha}) \\
    &= \mathbb{C}^{eq}(T_{env}) + \sum_{\alpha} [\mathbb{C}_{\alpha}^{src}(T_{\alpha}) - \mathbb{C}_{\alpha}^{src}(T_{env})].
\end{align}
Suppose that there is only one body. Then
\begin{align}
\label{eq:Cneqbodyenv}
    \mathbb{C}^{neq}(T_{env}, T_{body}) &= \mathbb{C}_{zp} + \mathbb{C}_{env}(T_{env}) + \mathbb{C}_{body}^{src}(T_{body}) \\
    &= \mathbb{C}^{eq}(T_{env}) + (\mathbb{C}_{body}^{src}(T_{body}) - \mathbb{C}_{body}^{src}(T_{env})) \nonumber
\end{align}
where
\begin{align}
    \mathbb{C}_{env}(T_{env}) &= -a(T_{env})\mathbb{G}_{body}\mathbb{G}_{0}^{-1\mathsf{A}}\mathbb{G}_{body}^{\dagger}, \\
    \mathbb{C}_{body}^{src}(T_{body}) &= a(T_{body}) \mathbb{G}_{body}\mathbb{V}_{body}^{\mathsf{A}}\mathbb{G}_{body}^{\dagger}.
\end{align}

We remark that it is possible to write the formulas using \(\mathbb{G}\) directly, but we choose to rewrite the quantities using the \(\mathbb{T}\) operator.
The bounds are found by maximizing an objective function with respect to possible scattering operator response, making an objective in terms of scattering operators more useful than one in terms of the total Green's function.
This is done by rewriting, using \(\mathbb{G}_{body} = \mathbb{G}_{0} + \mathbb{G}_{0}\mathbb{T}_{body}\mathbb{G}_{0}\), so that
\begin{align}
\label{eq:CenvToperator}
    \mathbb{C}_{env}(T_{env}) &= -a(T_{env})\mathbb{G}_{body}\mathbb{G}_{0}^{-1\mathsf{A}}\mathbb{G}_{body}^{\dagger} \nonumber\\
    &= a(T_{env})(\mathbb{I} + \mathbb{G}_{0}\mathbb{T}_{body})\mathbb{G}_{0}^{\mathsf{A}}(\mathbb{I} +\mathbb{T}_{body}^{\dagger} \mathbb{G}_{0}^{\dagger}).
\end{align}
\begin{align}
\label{eq:CbodyToperator}
    \mathbb{C}_{body}^{src}(T_{body}) &= a(T_{body}) \mathbb{G}_{body}\mathbb{V}_{body}^{\mathsf{A}}\mathbb{G}_{body}^{\dagger} \nonumber\\
    &= a(T_{body})\bigg((\mathbb{G}_{0} + \mathbb{G}_{0}\mathbb{T}_{body}\mathbb{G}_{0}) \nonumber \\
    & \qquad\qquad\qquad \mathbb{V}_{body}^{\mathsf{A}}(\mathbb{G}_{0} + \mathbb{G}_{0}\mathbb{T}_{body}\mathbb{G}_{0})^{\dagger}\bigg)
\end{align}

While the above equations are true, they do not lead directly to trace formulas. For this to occur, one must be able to extend integrals over a particular region to that of the entire space. A way to do this is to write the expressions so that the \(\mathbb{T}\) operator (or \(\mathbb{V}\)) appears on the right (or left) in integrands. Since the \(\mathbb{T}\) vanishes unless both spatial arguments are inside $\mathbb{V}$, one may then extend the integral over all space, resulting in a trace expression. Multiplying Eq.~\eqref{eq:CbodyToperator} on the right by \(\mathbb{G}_{0}^{-1\dagger}\) gives the correct form, but Eq.~\eqref{eq:CenvToperator} does not. A way around this is to note that (ignoring $\delta(\omega-\omega')$ factors)
\begin{align}
    \mathbb{C}_{body}^{src}(T) + \mathbb{C}_{env}(T) &= a(T) \mathbb{G}_{body}^{\mathsf{A}} \\
    &= a(T)(\mathbb{G}_{0} + \mathbb{G}_{0}\mathbb{T}_{body}\mathbb{G}_{0})^{\mathsf{A}},
\end{align}
which along with Eq.~\eqref{eq:CenvToperator} gives, after a few lines of algebraic manipulations,
\begin{align}
\label{eq:CbodysrcToperator}
    \mathbb{C}_{body}^{src}(T) &= a(T)\mathbb{G}_{0}\left(\mathbb{T}_{body}^{\mathsf{A}} - \mathbb{T}_{body}\mathbb{G}_{0}^{\mathsf{A}}\mathbb{T}_{body}^{\dagger} \right)\mathbb{G}_{0}^{\dagger},\\
    &\equiv a(T) \mathbb{R}_{body},
\end{align}
where in the last line we defined the radiation operator \(\mathbb{R}_{body},\) so called as it appears in
formulas involving a surface integral of the Poynting
vector~\cite{kruger_trace_formulae_for_nonequilibrium}. Note that this required that \(\mathbb{C}_{env}\)
and \(\mathbb{C}_{body}^{src}\) be evaluated at the same temperatures. This is where the utility of the
second line of Eq.~\eqref{eq:Cneqbodyenv} becomes apparent. Note also that
$\mathbb{C}_{body}^{src}\mathbb{G}_{0}^{-1\dagger}$ has a right-most $\mathbb{T}_{body}$ operator, so that
spatial integrals over the body $\int_{V_{body}} d\mathbf{r}(\dots)$ can be extended over all space if the
integrand depends on $\mathbb{C}_{body}^{src}\mathbb{G}_{0}^{-1\dagger},$ which it does for power and
force~\cite{kruger_trace_formulae_for_nonequilibrium} as well as for torque (see Eq.~\eqref{eq:torquetracewithC}).
The non-equilibrium portion of heat transfer, forces, and torques depends solely on \((\mathbb{C}_{body}^{src}(T_{body}) - \mathbb{C}_{body}^{src}(T_{env}))\).

\section{Evaluation of Green's function in spherical domains}

Once an origin has been specified, the Green's
function can be expanded in terms of the regular
spherical vector waves $\mathbf{RN}, \mathbf{RM}$ and in terms of the outgoing spherical vector waves
$\mathbf{N}, \mathbf{M}$ as ~\cite{tsang2000book}
\begin{align}
    \mathbb{G}_{0}(\mathbf{x}, \mathbf{y}) = -\frac{\delta(\mathbf{x} - \mathbf{y})}{k^{2}} \hat{x}\otimes\hat{y} + ik \sum_{J=1}^{\infty}\sum_{M=-J}^{J}(-1)^{M} \\
    \begin{cases}
    \mathbf{M}_{J,M}(k\mathbf{x})\mathbf{RM}_{J,-M}(k\mathbf{y}) + \mathbf{N}_{J,M}(k\mathbf{x})\mathbf{RN}_{J,-M}(k\mathbf{y}), x > y \\
    \mathbf{RM}_{J,M}(k\mathbf{x})\mathbf{M}_{J,-M}(k\mathbf{y}) + \mathbf{RN}_{J,M}(k\mathbf{x})\mathbf{N}_{J,-M}(k\mathbf{y}), x < y
    \end{cases} \nonumber
\end{align}
The asymmetric part has a spectral basis expansion as
\begin{align}
    \mathbb{G}_{0}^{\mathsf{A}}(\mathbf{x}, \mathbf{y}) = k\sum_{J,M}(-1)^{M} \Bigg( \mathbf{RM}_{J,M}(k\mathbf{x})\mathbf{RM}_{J,-M}(k\mathbf{y}) + \nonumber \\ \mathbf{RN}_{J,M}(k\mathbf{x})\mathbf{RN}_{J,-M}(k\mathbf{y})\Bigg)
\label{eq:G0ARgNRgM}
\end{align}
Explicitly,
\begin{align}
    \mathbf{RN}_{J,M}(\mathbf{y}) &= \frac{\sqrt{J(J+1)}}{y}j_{J}(y)\mathbf{A}_{JM}^{(3)} + \frac{1}{y}\frac{\partial(y j_{J}(y))}{\partial y}\mathbf{A}_{JM}^{(2)} \\
    \mathbf{RM}_{J,M}(\mathbf{y}) &= j_{J}(y)\mathbf{A}_{JM}^{(1)} \\
    \mathbf{N}_{J,M}(\mathbf{y}) &= \frac{\sqrt{J(J+1)}}{y}h_{J}^{(1)}(y)\mathbf{A}_{JM}^{(3)} + \frac{1}{y}\frac{\partial(y h_{J}^{(1)}(y))}{\partial y}\mathbf{A}_{JM}^{(2)} \\
    \mathbf{M}_{J,M}(\mathbf{y}) &= h_{J}^{(1)}(y)\mathbf{A}_{JM}^{(1)}
\end{align}
where $j_{J}(y)$ is the spherical Bessel function of order $J$ and $h_{J}^{(1)}(y)$ is the spherical Hankel function of order $J.$ 
\begin{align}
    \mathbf{A}_{JM}^{(1)}(\hat{\mathbf{r}}) &= \frac{1}{\sqrt{J(J+1)}} \nabla\times(\mathbf{r}Y_{JM}(\hat{\mathbf{r}})) \nonumber \\
    &= \frac{1}{\sqrt{J(J+1)}}\nabla Y_{JM}(\hat{\mathbf{r}})\times \mathbf{r}  \\
    \mathbf{A}_{JM}^{(2)}(\hat{\mathbf{r}}) &= \frac{1}{\sqrt{J(J+1)}}r\nabla Y_{JM}(\hat{\mathbf{r}}) \\
    \mathbf{A}_{JM}^{(3)}(\hat{\mathbf{r}}) &= \hat{\mathbf{r}} Y_{JM}(\hat{\mathbf{r}})
\end{align}
and the convention is such that
\begin{align}
    Y_{JM}(\theta, \phi) = \sqrt{\frac{2J+1}{4\pi}\frac{(J-M)!}{(J+M)!}}P_{J}^{M}(\cos{\theta})e^{im\phi}
\end{align}
where $P_{J}^{M}(z)$ is the associated Legendre polynomial.
The eigenvalues of $\mathbb{G}_{0}^{\mathsf{A}}$ are \cite{T_operator_bounds_angle_integrated}



\begin{align}
    \rho_{RM,J,M} &= \frac{\pi (kR)^{2}}{4k^{2}}\left(J_{J + \frac{1}{2}}^{2}(kR) - J_{J - \frac{1}{2}}(kR)J_{J + \frac{3}{2}}(kR)  \right)
\end{align}

\begin{align}
    \rho_{RN,J,M} &= \frac{\pi (kR)^{2}}{4k^{2}}\Bigg(
    \frac{J+1}{2J+1}\bigg(J_{J - \frac{1}{2}}^{2}(kR) \nonumber \\
    &- J_{J + \frac{1}{2}}(kR)J_{J 
    - \frac{3}{2}}(kR)\bigg) + \nonumber \\
    &\frac{J}{2J+1}\left(J_{J + \frac{3}{2}}^{2}(kR) - J_{J + \frac{1}{2}}(kR)J_{J + \frac{5}{2}}(kR)\right)
    \Bigg)
\end{align}
where $J_{\nu}$ is a Bessel function of the first kind of order $\nu.$

Note that $-i\frac{\partial}{\partial\phi}$ acting on the vector spherical harmonics does not simply introduce an overall factor of $m.$ This is because the coordinate vectors $\mathbf{e}_{r}, \mathbf{e}_{\theta}, \mathbf{e}_{\phi}$ depend on $\phi$ as well. More to the point, $-i\frac{\partial}{\partial\phi} = \hat{L}_{z}$ (math references usually have $\hbar = 1$) in the position space representation but the vector spherical harmonics are not defined to be eigenstates of $\hat{L}_{z}.$ That said, there does exist an operator, let us call it $\hat{J}_{z}$, equal to $\hat{J}_{z} = \hat{L}_{z} + \hat{S}_{z}$ for some operator $\hat{S}_{z}$ that cancels the terms introduced by the action of $\frac{\partial}{\partial\phi}$ on $\mathbf{e}_{r}, \mathbf{e}_{\theta}, \mathbf{e}_{\phi}$ such that the vector spherical harmonics are eigenfunctions of $\hat{J}_{z}$ with eigenvalues equal to the $m$ label. See Section \ref{sec:tensorsphericalharmonics} for more details.

\section{Tensor spherical harmonics}
\label{sec:tensorsphericalharmonics}

In this section, we summarize several salient mathematical identities and definitions from Ref.~\cite{khersonskii1988quantum} surrounding tensor spherical harmonics. The tensor spherical harmonics $Y_{JM}^{LS}(\theta,\phi)$ are, by definition, eigenfunctions of the operators $\hat{\mathbf{J}}^{2}, \hat{J}_{z}, \hat{\mathbf{L}}^{2},$ and $\hat{\mathbf{S}}^{2}$ where $\hat{\mathbf{L}}$ is the operator for the orbital angular momentum, $\hat{\mathbf{S}}$ is the operator for the spin, and $\hat{\mathbf{J}} = \hat{\mathbf{L}} + \hat{\mathbf{S}}$ is the operator for the total angular momentum. Explicitly,

\begin{align}
    \vecop{J}^{2}Y_{JM}^{LS}(\theta, \phi) &= J(J + 1)Y_{JM}^{LS}(\theta, \phi) \\
    \hat{J_{z}}Y_{JM}^{LS}(\theta, \phi) &= M Y_{JM}^{LS}(\theta, \phi) \\
    \vecop{L}^{2}Y_{JM}^{LS}(\theta, \phi) &= L(L + 1)Y_{JM}^{LS}(\theta, \phi) \\
    \vecop{S}^{2}Y_{JM}^{LS}(\theta, \phi) &= S(S + 1)Y_{JM}^{LS}(\theta, \phi).
\end{align}

Note that the units are such that $\hbar = 1.$ The interpretation within physics is that the tensor spherical harmonics may be used in the expansion of the angular distribution and polarization of spin-$S$ particles. The tensor spherical harmonics are states with definite total angular momentum $J$, definite projection $M$ along an axis (chosen to be the $z$ axis), and definite orbital angular momentum $L.$ For $S=0,$ these are just the spherical harmonics, often written $Y_{lm}(\theta, \phi)$ or $Y_{l}^{m}(\theta, \phi).$ The $S = \frac{1}{2}$ functions are sometimes called spinor spherical harmonics. The $S = 1$ states are the vector spherical harmonics, etc. (The transformation properties of the tensor spherical harmonics under a rotation of the coordinate system are determined by $J,$ and not $L$ or $S$ so calling them spinor or vector spherical harmonics is a bit of a misnomer from this point of view.)

The tensor spherical harmonics may be constructed from the spherical harmonics, let us label them $Y_{LM}(\theta, \phi)$ (instead of $Y_{lm}(\theta, \phi)$ of $Y_{l}^{m}(\theta, \phi)).$ The expansion is
\begin{align}
\label{eq:tensorsphericalharmonicexpansion}
    Y_{JM}^{LS}(\theta, \phi) = \sum_{M, \sigma} C_{LMS\sigma}^{JM} Y_{LM}(\theta, \phi) \chi_{S\sigma}.
\end{align}
This follows from the fact that any total angular momentum basis $\ket{jm ; l\sigma}$ can be expanded in terms of the direct product basis $\ket{l m_{l} ; \sigma m_{\sigma}}\equiv \ket{l m_{l}}\otimes \ket{\sigma m_{\sigma}}$ and the Clebsch-Gordon coefficients
\begin{align}
    \ket{jm; l\sigma} = \sum_{m_{l} = - l}^{l}\sum_{m_{\sigma}=-\sigma}^{\sigma} \braket{l m_{l}; \sigma m_{\sigma}}{jm; l \sigma} \ket{lm_{l}; \sigma m_{\sigma}}
\end{align}
In the coordinate representation $\vecop{L}$ is represented as a differential operator $\vecop{L} = -i\mathbf{r}\times \nabla.$ Using the coordinate representation and the fact that, by definition, the spherical harmonics $Y_{lm}(\theta, \phi)$ satisfy
\begin{align}
    \vecop{L}^{2}Y_{lm}(\theta, \phi) &= l(l+1) Y_{lm}(\theta, \phi) \\
    \hat{L}_{z} Y_{lm}(\theta, \phi) &= m Y_{lm}(\theta, \phi)
\end{align}
gives the coordinate representation claimed in Eq.~\eqref{eq:tensorsphericalharmonicexpansion}. The indices $J$ and $S$ are integer or half-integer nonnegative numbers. $L$ is a nonnegative integer. For a fixed value of $J$ and $S,$ then $L$ can only take on values from $|J-S|, |J-S| + 1, \dots, J+S-1, J+S.$ For a given value of $J,$ then $M$ can only take on values from $-J, -J + 1, \dots, J-1, J.$

For fixed values of $J, M, L, S$ the tensor spherical harmonics are function of $\theta, \phi,$ and $\xi,$ where $\xi$ is the spin variable. The polar angles are $\theta, \phi$ take on values $0 \leq \theta \leq \pi$ and $0 \leq \phi \leq 2\pi.$ To be precise, one should write $Y_{JM}^{LS}(\theta, \phi, \xi)$ but the dependence on the spin variable is usually not explicitly mentioned. The reason for this is because of the next step: Represent $Y_{JM}^{LS}(\theta, \phi)$ as a column matrix (that is, a column vector) of length $2S + 1.$ Therefore, the spin variable now refers to a particular component of the column vector, and summation of spin variables has the interpretation of matrix multiplication. Using the matrix notation, the following orthogonality relation holds:
\begin{align}
    \sum_{M} (Y_{LM}^{L',S}(\theta, \phi))^{*T}\cdot Y_{JM}^{LS}(\theta, \phi) = 0, \text{ if } L' \neq L.
\end{align}

\subsection{Vector spherical harmonics}

The vector spherical harmonics are defined as the tensor spherical harmonics with $S=1;$ they are $Y_{JM}^{L,1}(\theta, \phi).$ Using the vector notation,

\begin{align}
    Y_{JM}^{L,1}(\theta, \phi) = \sum_{M, \sigma} C_{LM1\sigma}^{JM} Y_{LM}(\theta, \phi) \mathbf{e}_{\sigma},
\end{align}
where $\mathbf{e}_{\sigma}$ are a covariant spherical basis vector. That is,
\begin{subequations}
 \label{eq:sphericalbasisintermsofcartesianbasis}
\begin{align}
    \mathbf{e}_{+1} &= -\frac{1}{\sqrt{2}}(\mathbf{e}_{x} + i\mathbf{e}_{y}), \\
    \mathbf{e}_{0} &= \mathbf{e}_{z}, \\
    \mathbf{e}_{-1} &= \frac{1}{\sqrt{2}}(\mathbf{e}_{x} - i\mathbf{e}_{y}).
\end{align}
\end{subequations}
For a fixed value of $J,$ the possible values of $L$ are $J-1, J, J+1,$ with the exception of $J=0,$ where only $L = 1$ is allowed.

While the covariant spherical basis vectors are nice from a mathematical theoretical point of view and lead to cleaner transformation properties of their components under rotations of coordinate systems, other bases are possible. The change of basis formulas from covariant spherical basis to Cartesian or polar coordinates are straightforward (see below).

Given this introduction to tensor spherical harmonics, a valid question is how does this relate to $\mathbf{A}_{JM}^{(1)}, \mathbf{A}_{JM}^{(2)}, \mathbf{A}_{JM}^{(3)}$ which are also called vector spherical harmonics. Note that $Y_{JM}^{LS}$ are, by definition, states of definite orbital angular momentum. In the context of radiation settings in electromagnetism, it can be convenient to work in a basis that separates longitudinal and transverse waves. It turns out that
\begin{align}
    \mathbf{A}_{JM}^{(1)}(\hat{\mathbf{r}}) &= -iY_{JM}^{J, 1}(\theta, \phi), \\
    \mathbf{A}_{JM}^{(2)}(\hat{\mathbf{r}}) &= \sqrt{\frac{J + 1}{2J + 1}} Y_{JM}^{J-1, 1}(\theta, \phi) + \sqrt{\frac{J}{2J + 1}} Y_{JM}^{J+1, 1}(\theta, \phi), \\
    \mathbf{A}_{JM}^{(3)}(\hat{\mathbf{r}}) &= \sqrt{\frac{J}{2J + 1}} Y_{JM}^{J-1, 1}(\theta, \phi) - \sqrt{\frac{J + 1}{2J + 1}} Y_{JM}^{J+1, 1}(\theta, \phi)
\end{align}
are the needed combinations of the tensor spherical harmonics (up to overall constant complex factors) for the decomposition into transverse and longitudinal waves. $\mathbf{A}_{JM}^{(3)}$ are longitudinal waves. $\mathbf{A}_{JM}^{(1)}$ and $\mathbf{A}_{JM}^{(2)}$ are transverse waves, sometimes called magnetic and electric multipoles, respectively. See Chapter 7 of Ref.~\cite{khersonskii1988quantum} for more details. Ref.~\cite{totalangularmomentumwavesforfields} explicitly works out the divergence of the tensor spherical waves and shows how to use the expressions for the divergence to construct linear combinations of the tensor spherical waves that are longitudinal and transverse.
This process is invertible, namely,
\begin{align}
     Y_{JM}^{J, 1}(\theta, \phi) &= i\mathbf{A}_{JM}^{(1)}(\hat{\mathbf{r}}), \\
    Y_{J,M}^{J+1,1}(\theta,\phi) &= \sqrt{\frac{J}{2J + 1}} \mathbf{A}_{JM}^{(2)}(\hat{\mathbf{r}}) - \sqrt{\frac{J+1}{2J + 1}} \mathbf{A}_{JM}^{(3)}(\hat{\mathbf{r}}), \\
    Y_{JM}^{J-1,1}(\theta,\phi) &= \sqrt{\frac{J+1}{2J + 1}} \mathbf{A}_{JM}^{(2)}(\hat{\mathbf{r}}) + \sqrt{\frac{J}{2J + 1}} \mathbf{A}_{JM}^{(3)}(\hat{\mathbf{r}})
\end{align}
so $\mathbf{A}_{JM}^{(1)}, \mathbf{A}_{JM}^{(2)}, \mathbf{A}_{JM}^{(3)}$ also constitute a complete orthonormal vector set for the range $0 \leq \theta \leq \pi,$ $0 \leq \phi \leq 2\pi.$ Their longitudinal and transverse orientations relative to $\hat{\mathbf{r}}$ makes them a convenient basis to use in radiation settings in electromagnetism. 

\subsection{The $\hat{S}_{z}$ Operator}

In the covariant spherical basis \cite{khersonskii1988quantum}, 
\begin{align}
    \hat{S}_{z} =
    \begin{bmatrix}
        1 & 0 & 0 \\
        0 & 0 & 0 \\
        0 & 0 & -1
    \end{bmatrix}.
\end{align}
From Eq.~\eqref{eq:sphericalbasisintermsofcartesianbasis}, it follows that the change-of-basis operator from $\{\mathbf{e}_{+1}, \mathbf{e}_{0}, \mathbf{e}_{-1}\}$ to $\{\mathbf{e}_{x}, \mathbf{e}_{y}, \mathbf{e}_{z}\}$ is
\begin{align}
    M(x,y,z \leftarrow +1,0,-1) =
    \begin{bmatrix}
        -\frac{1}{\sqrt{2}} & 0 & \frac{1}{\sqrt{2}} \\
        -\frac{i}{\sqrt{2}} & 0 & -\frac{i}{\sqrt{2}} \\
        0 & 1 & 0
    \end{bmatrix}.
\end{align}
Likewise, the change-of-basis operator from $\{\mathbf{e}_{+1}, \mathbf{e}_{0}, \mathbf{e}_{-1}\}$ to $\{\mathbf{e}_{r}, \mathbf{e}_{\theta}, \mathbf{e}_{\phi}\}$ is
\begin{align}
    M(r,\theta,\phi \leftarrow +1,0,-1) =
    \begin{bmatrix}
        -\frac{\sin\theta}{\sqrt{2}}e^{i\phi} & \cos\theta & \frac{\sin\theta}{\sqrt{2}}e^{-i\phi} \\
        -\frac{\cos\theta}{\sqrt{2}}e^{i\phi} & -\sin\theta & \frac{\cos\theta}{\sqrt{2}}e^{-i\phi} \\
        -\frac{i}{\sqrt{2}}e^{i\phi} & 0 & -\frac{i}{\sqrt{2}}e^{-i\phi}
    \end{bmatrix}.
\end{align}
It follows that in the polar coordinate basis $\{\mathbf{e}_{r}, \mathbf{e}_{\theta}, \mathbf{e}_{\phi}\}$
\begin{align}
    \hat{S}_{z} =
    \begin{bmatrix}
        0 & 0 & -i\sin\theta \\
        0 & 0 & -i\cos\theta \\
        i\sin\theta & i\cos\theta & 0
    \end{bmatrix}.
\end{align}
From these change-of-basis operators, it follows that in the Cartesian coordinate basis $\{\mathbf{e}_{x}, \mathbf{e}_{y}, \mathbf{e}_{z}\}$
\begin{align}
    \hat{S}_{z}
    =
    \begin{bmatrix}
        0 & -i & 0 \\
        i & 0 & 0 \\
        0 & 0 & 0
    \end{bmatrix}. 
\end{align}

\subsection{Action of $\hat{L}_{z}$ and $\hat{S}_{z}$ on $Y_{JM}^{LS}$}

Let $\Phi(r)$ be an arbitrary function of $r = |\mathbf{r}|.$ From Ref.~\cite{khersonskii1988quantum}, the following holds:

\begin{widetext}
    \begin{align}
        \hat{L}_{\mu}\{\Phi(r)Y_{JM}^{LS}(\theta, \phi)\} &= \Phi(r)\hat{L}_{\mu}\{Y_{JM}^{LS}(\theta,\phi)\} \nonumber\\
        &= (-1)^{J+L+S+1}\Phi(r)\sqrt{(2J+1)L(L+1)(2L+1)}\sum_{J'}
        \begin{Bmatrix}
        J & J' & 1 \\
        L & L & S
     \end{Bmatrix}
        C_{JM1\mu}^{J'M+\mu}Y_{J'M+\mu}^{LS}(\theta,\phi).
    \end{align}
    \begin{align}
        \hat{S}_{\mu}\{\Phi(r)Y_{JM}^{LS}(\theta, \phi)\} &= \Phi(r)\hat{S}_{\mu}\{Y_{JM}^{LS}(\theta,\phi)\} \nonumber\\
        &= (-1)^{L+1}\Phi(r)\sqrt{(2J+1)S(S+1)(2S+1)}\sum_{J'}
        (-1)^{J'+S}
        \begin{Bmatrix}
        J & J' & 1 \\
        S & S & L
     \end{Bmatrix}
        C_{JM1\mu}^{J'M+\mu}Y_{J'M+\mu}^{LS}(\theta,\phi).
    \end{align}

When $\mu = 0,$ the spherical coordinate components are equal to the Cartesian coordinate components. Namely,

    \begin{align}
        \hat{L}_{z}\{\Phi(r)Y_{JM}^{LS}(\theta, \phi)\} &= \Phi(r)\hat{L}_{z}\{Y_{JM}^{LS}(\theta,\phi)\} \nonumber\\
        &= (-1)^{J+L+S+1}\Phi(r)\sqrt{(2J+1)L(L+1)(2L+1)}\sum_{J'}
        \begin{Bmatrix}
        J & J' & 1 \\
        L & L & S
     \end{Bmatrix}
        C_{JM10}^{J'M}Y_{J'M}^{LS}(\theta,\phi).
        \label{eq:LzYJMLS}
    \end{align}
    \begin{align}
        \hat{S}_{z}\{\Phi(r)Y_{JM}^{LS}(\theta, \phi)\} &= \Phi(r)\hat{S}_{z}\{Y_{JM}^{LS}(\theta,\phi)\} \nonumber\\
        &= (-1)^{L+1}\Phi(r)\sqrt{(2J+1)S(S+1)(2S+1)}\sum_{J'}
        (-1)^{J'+S}
        \begin{Bmatrix}
        J & J' & 1 \\
        S & S & L
     \end{Bmatrix}
        C_{JM10}^{J'M}Y_{J'M}^{LS}(\theta,\phi).
    \end{align}
\end{widetext}

\subsection{Action of $\hat{S}_{z}$ on $\mathbf{A}_{JM}^{(1)}, \mathbf{A}_{JM}^{(2)}, \mathbf{A}_{JM}^{(3)}$}

Using the results of the previous parts,

\begin{widetext}
    
\begin{align}
    \hat{S}_{z}\mathbf{A}_{JM}^{(1)} &= -i(-1)^{J+1}\sqrt{(2J+1)S(S+1)(2S+1)}\sum_{J'}
        (-1)^{J'+1}
        \begin{Bmatrix}
        J & J' & 1 \\
        1 & 1 & J
        \end{Bmatrix}
        C_{JM10}^{J'M}Y_{J'M}^{J,1} \\
        &= -i(-1)^{J+1}\sqrt{(2J+1)S(S+1)(2S+1)}\bigg( 
        (-1)^{J}
        \begin{Bmatrix}
        J & J-1 & 1 \\
        1 & 1 & J
        \end{Bmatrix}C^{J-1,M}_{JM10}Y^{J,1}_{J-1,M} \nonumber \\
        &+
        (-1)^{J+1}
        \begin{Bmatrix}
        J & J & 1 \\
        1 & 1 & J
        \end{Bmatrix}C^{J,M}_{JM10}Y^{J,1}_{J,M}
        \nonumber \\
        &+
        (-1)^{J}
        \begin{Bmatrix}
        J & J+1 & 1 \\
        1 & 1 & J
        \end{Bmatrix}C^{J+1,M}_{JM10}Y^{J,1}_{J+1,M} \bigg)\nonumber \\
        &= 
        -i(-1)^{J+1}\sqrt{(2J+1)S(S+1)(2S+1)}\bigg( \nonumber \\
        &
        (-1)^{J}
        \begin{Bmatrix}
        J & J-1 & 1 \\
        1 & 1 & J
        \end{Bmatrix}C^{J-1,M}_{JM10}\sqrt{\frac{J-1}{2J - 1}}\mathbf{A}_{J-1,M}^{(2)}
        -
        (-1)^{J}
        \begin{Bmatrix}
        J & J-1 & 1 \\
        1 & 1 & J
        \end{Bmatrix}C^{J-1,M}_{JM10}\sqrt{\frac{J}{2J - 1}}\mathbf{A}_{J-1,M}^{(3)}\nonumber \\
        &+
        (-1)^{J+1}
        \begin{Bmatrix}
        J & J & 1 \\
        1 & 1 & J
        \end{Bmatrix}C^{J,M}_{JM10}i\mathbf{A}_{JM}^{(1)}
         \nonumber \\
        &+
        (-1)^{J}
        \begin{Bmatrix}
        J & J+1 & 1 \\
        1 & 1 & J
        \end{Bmatrix}C^{J+1,M}_{JM10}\sqrt{\frac{J+2}{2J+3}}\mathbf{A}_{J+1,M}^{(2)}
        +
        (-1)^{J}
        \begin{Bmatrix}
        J & J+1 & 1 \\
        1 & 1 & J
        \end{Bmatrix}C^{J+1,M}_{JM10}\sqrt{\frac{J+1}{2J+3}}\mathbf{A}_{J+1,M}^{(3)}
        \bigg) \nonumber
\end{align}

\begin{align}
    \hat{S}_{z}\mathbf{A}_{JM}^{(2)} &= \sqrt{\frac{J+1}{2J+1}}(-1)^{J}\sqrt{(2J+1)S(S+1)(2S+1)}\sum_{J'}
        (-1)^{J'+1}
        \begin{Bmatrix}
        J & J' & 1 \\
        1 & 1 & J-1
        \end{Bmatrix}
        C_{JM10}^{J'M}Y_{J'M}^{J-1,1}  \\
        &+ \sqrt{\frac{J}{2J+1}}(-1)^{J}\sqrt{(2J+1)S(S+1)(2S+1)}\sum_{J'}
        (-1)^{J'+1}
        \begin{Bmatrix}
        J & J' & 1 \\
        1 & 1 & J+1
        \end{Bmatrix}
        C_{JM10}^{J'M}Y_{J'M}^{J+1,1} \nonumber \\
        &= (-1)^{J}\sqrt{\frac{J+1}{2J+1}}\sqrt{(2J+1)S(S+1)(2S+1)}\bigg(  \nonumber \\
        &
        (-1)^{J}
        \begin{Bmatrix}
        J & J-1 & 1 \\
        1 & 1 & J-1
        \end{Bmatrix}C^{J-1,M}_{JM10}Y^{J-1,1}_{J-1,M} \nonumber \\
        &+
        (-1)^{J+1}
        \begin{Bmatrix}
        J & J & 1 \\
        1 & 1 & J - 1
        \end{Bmatrix}C^{J,M}_{JM10}Y^{J-1,1}_{J,M}
        \bigg) \nonumber \\
        &
        +(-1)^{J}\sqrt{\frac{J}{2J+1}}\sqrt{(2J+1)S(S+1)(2S+1)}\bigg( \nonumber \\
        &
        (-1)^{J+1}
        \begin{Bmatrix}
        J & J  & 1 \\
        1 & 1 & J+1
        \end{Bmatrix}
        C_{JM10}^{J,M}Y_{JM}^{J+1,1} \nonumber \\
        &
        +
        (-1)^{J}
        \begin{Bmatrix}
        J & J+1  & 1 \\
        1 & 1 & J+1
        \end{Bmatrix}
        C_{JM10}^{J+1,M}Y_{J+1,M}^{J+1,1} \bigg)\nonumber \\
        &= (-1)^{J}\sqrt{\frac{J+1}{2J+1}}\sqrt{(2J+1)S(S+1)(2S+1)}\bigg( \nonumber \\
        &
        (-1)^{J}
        \begin{Bmatrix}
        J & J-1 & 1 \\
        1 & 1 & J-1
        \end{Bmatrix}C^{J-1,M}_{JM10} i\mathbf{A}_{J-1,M}^{(1)}
        \nonumber \\
        &+
        (-1)^{J+1}
        \begin{Bmatrix}
        J & J & 1 \\
        1 & 1 & J - 1
        \end{Bmatrix}C^{J,M}_{JM10} \sqrt{\frac{J+1}{2J+1}}\mathbf{A}_{J,M}^{(2)}
        +
        (-1)^{J+1}
        \begin{Bmatrix}
        J & J & 1 \\
        1 & 1 & J - 1
        \end{Bmatrix}C^{J,M}_{JM10} \sqrt{\frac{J}{2J+1}}\mathbf{A}_{J,M}^{(3)}
        \bigg) \nonumber \\
        &
        +(-1)^{J}\sqrt{\frac{J}{2J+1}}\sqrt{(2J+1)S(S+1)(2S+1)}\bigg( \nonumber \\
        &
        (-1)^{J+1}
        \begin{Bmatrix}
        J & J  & 1 \\
        1 & 1 & J+1
        \end{Bmatrix}
        C_{JM10}^{J,M}\sqrt{\frac{J}{2J+1}}\mathbf{A}_{JM}^{(2)} -
        (-1)^{J+1}
        \begin{Bmatrix}
        J & J  & 1 \\
        1 & 1 & J+1
        \end{Bmatrix}
        C_{JM10}^{J,M}\sqrt{\frac{J+1}{2J+1}}\mathbf{A}_{JM}^{(3)}\nonumber \\
        &+
        (-1)^{J}
        \begin{Bmatrix}
        J & J+1  & 1 \\
        1 & 1 & J+1
        \end{Bmatrix}
        C_{JM10}^{J+1,M}i\mathbf{A}_{J+1,M}^{(1)} \bigg)\nonumber 
\end{align}

\begin{align}
    \hat{S}_{z}\mathbf{A}_{JM}^{(3)} &= \sqrt{\frac{J}{2J+1}}(-1)^{J}\sqrt{(2J+1)S(S+1)(2S+1)}\sum_{J'}
        (-1)^{J'+1}
        \begin{Bmatrix}
        J & J' & 1 \\
        1 & 1 & J-1
        \end{Bmatrix}
        C_{JM10}^{J'M}Y_{J'M}^{J-1,1} \\
        &- \sqrt{\frac{J+1}{2J+1}}(-1)^{J}\sqrt{(2J+1)S(S+1)(2S+1)}\sum_{J'}
        (-1)^{J'+1}
        \begin{Bmatrix}
        J & J' & 1 \\
        1 & 1 & J+1
        \end{Bmatrix}
        C_{JM10}^{J'M}Y_{J'M}^{J+1,1} \nonumber \\
        &= (-1)^{J}\sqrt{\frac{J}{2J+1}}\sqrt{(2J+1)S(S+1)(2S+1)}\bigg(  \nonumber \\
        &
        (-1)^{J}
        \begin{Bmatrix}
        J & J-1 & 1 \\
        1 & 1 & J-1
        \end{Bmatrix}C^{J-1,M}_{JM10}Y^{J-1,1}_{J-1,M} \nonumber \\
        &+
        (-1)^{J+1}
        \begin{Bmatrix}
        J & J & 1 \\
        1 & 1 & J - 1
        \end{Bmatrix}C^{J,M}_{JM10}Y^{J-1,1}_{J,M}
        \bigg) \nonumber \\
        &
        -(-1)^{J}\sqrt{\frac{J+1}{2J+1}}\sqrt{(2J+1)S(S+1)(2S+1)}\bigg( \nonumber \\
        &
        (-1)^{J+1}
        \begin{Bmatrix}
        J & J  & 1 \\
        1 & 1 & J+1
        \end{Bmatrix}
        C_{JM10}^{J,M}Y_{JM}^{J+1,1} \nonumber \\
        &
        +
        (-1)^{J}
        \begin{Bmatrix}
        J & J+1  & 1 \\
        1 & 1 & J+1
        \end{Bmatrix}
        C_{JM10}^{J+1,M}Y_{J+1,M}^{J+1,1} \bigg)\nonumber \\
        &= (-1)^{J}\sqrt{\frac{J}{2J+1}}\sqrt{(2J+1)S(S+1)(2S+1)}\bigg(  \nonumber \\
        &
        (-1)^{J}
        \begin{Bmatrix}
        J & J-1 & 1 \\
        1 & 1 & J-1
        \end{Bmatrix}C^{J-1,M}_{JM10} i\mathbf{A}_{J-1,M}^{(1)}
        \nonumber \\
        &+
        (-1)^{J+1}
        \begin{Bmatrix}
        J & J & 1 \\
        1 & 1 & J - 1
        \end{Bmatrix}C^{J,M}_{JM10} \sqrt{\frac{J+1}{2J+1}}\mathbf{A}_{J,M}^{(2)}
        +
        (-1)^{J+1}
        \begin{Bmatrix}
        J & J & 1 \\
        1 & 1 & J - 1
        \end{Bmatrix}C^{J,M}_{JM10} \sqrt{\frac{J}{2J+1}}\mathbf{A}_{J,M}^{(3)}
        \bigg) \nonumber \\
        &
        -(-1)^{J}\sqrt{\frac{J+1}{2J+1}}\sqrt{(2J+1)S(S+1)(2S+1)}\bigg( \nonumber\\
        &
        (-1)^{J+1}
        \begin{Bmatrix}
        J & J  & 1 \\
        1 & 1 & J+1
        \end{Bmatrix}
        C_{JM10}^{J,M}\sqrt{\frac{J}{2J+1}}\mathbf{A}_{JM}^{(2)} -
        (-1)^{J+1}
        \begin{Bmatrix}
        J & J  & 1 \\
        1 & 1 & J+1
        \end{Bmatrix}
        C_{JM10}^{J,M}\sqrt{\frac{J+1}{2J+1}}\mathbf{A}_{JM}^{(3)}\nonumber \\
        &+
        (-1)^{J}
        \begin{Bmatrix}
        J & J+1  & 1 \\
        1 & 1 & J+1
        \end{Bmatrix}
        C_{JM10}^{J+1,M}i\mathbf{A}_{J+1,M}^{(1)} \bigg)\nonumber 
\end{align}
Analogous expressions for the action of $\hat{L}_{z}$ on $\mathbf{A}_{JM}^{(1)}, \mathbf{A}_{JM}^{(2)}, \mathbf{A}_{JM}^{(3)}$ can be derived starting from Eq.~\eqref{eq:LzYJMLS} and following steps similar to the above work. One can also start from $\hat{L}_{z}\mathbf{A}_{JM} = \hat{J}_{z}\mathbf{A}_{JM} - \hat{S}_{z}\mathbf{A}_{JM} = M\mathbf{A}_{JM} - \hat{S}_{z}\mathbf{A}_{JM}$ and then use the expressions just derived.
\end{widetext}

\section{Contibutions to $\Phi_{J}$ in the small $R$ limit}
\begin{widetext}
Let $R$ be a measure of the size of the compact body. In this section, we show that in the small $R$ limit
the contributions from the $\hat{L}_{z}$ terms vanish exactly to lowest order in $R$ whereas they do not, in general, vanish from the $\hat{S}_{z}$ terms. This supports the intuitive semi-classical picture that the spin contributions dominate in the quasistatic regime.
In the small $R$ limit, the $\mathbf{RN}_{JM}$ with $J=1$ terms dominate in $\Phi_{J}.$ In this limit (using $\lim_{x\to 0} \frac{\partial(xj_{J}(x))}{\partial x} \approx (J+1)j_{J}(x)$) one finds
\begin{align}
    \hat{J}_{z}\mathbb{G}_{0}^{\mathsf{A}}(\mathbf{x}, \mathbf{y}) &\approx \sum_{M=-1,0,1}M
    \mathbf{RN}_{1,M}(\mathbf{x})\mathbf{RN}_{1,M}^{*}(\mathbf{y})\\
    &\approx \sum_{M=-1,0,1} \frac{j_{1}(x)j_{1}(y)}{xy}
    \begin{bmatrix}
    \mathbf{A}_{1M}^{(1)}(\mathbf{x}) \\
    \mathbf{A}_{1M}^{(2)}(\mathbf{x}) \\
    \mathbf{A}_{1M}^{(3)}(\mathbf{x})
    \end{bmatrix}^{T}
    \mathcal{A}_{M}
    \begin{bmatrix}
    \mathbf{A}_{1M}^{(1)*}(\mathbf{y}) \\
    \mathbf{A}_{1M}^{(2)*}(\mathbf{y}) \\
    \mathbf{A}_{1M}^{(3)*}(\mathbf{y})
    \end{bmatrix}
    \end{align}
where
\begin{align}
    \mathcal{A}_{-1} =
    \begin{bmatrix}
    0 & 0 & 0 \\
    0 & -4 & -2\sqrt{2} \\
    0 & -2\sqrt{2} & -2
    \end{bmatrix},
    \quad
    \mathcal{A}_{0} =
    \begin{bmatrix}
    0 & 0 & 0 \\
    0 & 0 & 0 \\
    0 & 0 & 0
    \end{bmatrix},
    \quad
    \mathcal{A}_{1} =
    \begin{bmatrix}
    0 & 0 & 0 \\
    0 & 4 & 2\sqrt{2} \\
    0 & 2\sqrt{2} & 2
    \end{bmatrix}.
\end{align}
In this small $R$ limit, $\Phi_{J} \approx -\frac{2}{\pi}\text{Tr}[\hat{J}_{z}\mathbb{G}_{0}^{\mathsf{A}}\mathbb{T}^{\mathsf{A}}] = -\frac{2}{\pi}\text{Tr}[\hat{L}_{z}\mathbb{G}_{0}^{\mathsf{A}}\mathbb{T}^{\mathsf{A}}] - \frac{2}{\pi}\text{Tr}[\hat{S}_{z}\mathbb{G}_{0}^{\mathsf{A}}\mathbb{T}^{\mathsf{A}}].$ From the expressions in the previous sections for the action of $\hat{S}_{z}$ on the VSHs, one can extract scaling behaviors of the $\hat{S}_{z}$ and $\hat{L}_{z}$ contributions. To do this, one must rewrite $\mathbb{G}_{0}^{\mathsf{A}}$ as an outerproduct of the vectors $\mathbf{A}_{JM}^{(1)}, \mathbf{A}_{JM}^{(2)}, \mathbf{A}_{JM}^{(3)}$ and then act with $\hat{S}_{z}.$ After plugging in the definition of $\mathbf{RN}_{JM}$ and $\mathbf{RM}_{JM}$ one finds
    \begin{align}
    \hat{S}_{z}\mathbb{G}_{0}^{\mathsf{A}}(\mathbf{x}, \mathbf{y}) =
    &\sum_{JM} j_{J}(x)j_{J}(y) \hat{S}_{z}\mathbf{A}_{JM}^{(1)}(\mathbf{x})\mathbf{A}_{JM}^{(1)*}(\mathbf{y})
    +
    \sum_{JM} \frac{J(J+1)}{xy}j_{J}(x)j_{J}(y) \hat{S}_{z}\mathbf{A}_{JM}^{(3)}(\mathbf{x})\mathbf{A}_{JM}^{(3)*}(\mathbf{y}) \nonumber \\
    &+
    \sum_{JM} \frac{\sqrt{J(J+1)}}{xy}j_{J}(x)\frac{\partial(yj_{J}(y))}{\partial y} \hat{S}_{z}\mathbf{A}_{JM}^{(3)}(\mathbf{x})\mathbf{A}_{JM}^{(2)*}(\mathbf{y}) \nonumber \\
    &+
    \sum_{JM} \frac{\sqrt{J(J+1)}}{xy}j_{J}(y)\frac{\partial(xj_{J}(x))}{\partial x} \hat{S}_{z}\mathbf{A}_{JM}^{(2)}(\mathbf{x})\mathbf{A}_{JM}^{(3)*}(\mathbf{y}) \nonumber \\
    &+\sum_{JM} \frac{1}{xy}\frac{\partial(xj_{J}(x))}{\partial x}\frac{\partial(yj_{J}(y))}{\partial y} \hat{S}_{z}\mathbf{A}_{JM}^{(2)}(\mathbf{x})\mathbf{A}_{JM}^{(2)*}(\mathbf{y}).
    \label{eq:SzG0AsmallR}
    \end{align}
In the limit that the object size $R$ approaches 0, then the $x$ and $y$ arguments in the above expression will also approach 0 when evaluating the trace over the object. But $\lim_{x\to 0} j_{J}(x) = \frac{2^{J}}{(2J+1)!}x^{J}$ and $\lim_{x\to 0} \frac{\partial(xj_{J}(x))}{\partial x} \approx (J+1)j_{J}(x)$ so that
    \begin{align}
    \hat{S}_{z}\mathbb{G}_{0}^{\mathsf{A}}(\mathbf{x}, \mathbf{y}) =
    &\sum_{JM} j_{J}(x)j_{J}(y) \hat{S}_{z}\mathbf{A}_{JM}^{(1)}(\mathbf{x})\mathbf{A}_{JM}^{(1)*}(\mathbf{y})
    +
    \sum_{JM} \frac{J(J+1)}{xy}j_{J}(x)j_{J}(y) \hat{S}_{z}\mathbf{A}_{JM}^{(3)}(\mathbf{x})\mathbf{A}_{JM}^{(3)*}(\mathbf{y}) \nonumber \\
    &+
    \sum_{JM} \frac{\sqrt{J(J+1)}(J+1)}{xy}j_{J}(x)j_{J}(y) \hat{S}_{z}\mathbf{A}_{JM}^{(3)}(\mathbf{x})\mathbf{A}_{JM}^{(2)*}(\mathbf{y}) \nonumber \\
    &+
    \sum_{JM} \frac{\sqrt{J(J+1)}(J+1)}{xy}j_{J}(y)j_{J}(x)) \hat{S}_{z}\mathbf{A}_{JM}^{(2)}(\mathbf{x})\mathbf{A}_{JM}^{(3)*}(\mathbf{y}) \nonumber \\
    &+\sum_{JM} \frac{(J+1)^{2}}{xy}j_{J}(x)j_{J}(y) \hat{S}_{z}\mathbf{A}_{JM}^{(2)}(\mathbf{x})\mathbf{A}_{JM}^{(2)*}(\mathbf{y}).
    \end{align}
This then allows one to extract the dominate terms in the small $R$ limit. The lowest order terms appear in the $J=1$ terms. In particular, the $j_{J}(x)j_{J}(y) \sim R^{2J}$ and $j_{J}(x)j_{J}(y)/(xy) \sim R^{2J-2}.$ Ultimately, one needs the trace and there is also $\mathbb{T}^{\mathsf{A}}$ which contains a delta function, so the overall scaling in the final trace gets an additional $R^{3}$ factor. The smallest in $R$ terms come from the $J=1$ terms in $j_{J}(x)j_{J}(y)/(xy),$ which scales like $R^{2(1)-2 +3} = R^{3}$ in $\Phi_{S}.$ Plugging in the expressions for $\hat{S}_{z}\mathbf{A}_{JM}^{(k)}$ for $k=1,2,3$ and simplifying one finds that 
    \begin{align}
    \hat{S}_{z}\mathbb{G}_{0}^{\mathsf{A}}(\mathbf{x}, \mathbf{y}) =
    &\sum_{M=-1,0,1} \frac{j_{1}(x)j_{1}(y)}{xy}
    \begin{bmatrix}
    \mathbf{A}_{1M}^{(1)}(\mathbf{x}) \\
    \mathbf{A}_{1M}^{(2)}(\mathbf{x}) \\
    \mathbf{A}_{1M}^{(3)}(\mathbf{x})
    \end{bmatrix}^{T}
    \mathcal{A}_{M}
    \begin{bmatrix}
    \mathbf{A}_{1M}^{(1)*}(\mathbf{y}) \\
    \mathbf{A}_{1M}^{(2)*}(\mathbf{y}) \\
    \mathbf{A}_{1M}^{(3)*}(\mathbf{y})
    \end{bmatrix}
    + \mathcal{O}(R^{1})
    \end{align}
where, again,
\begin{align}
    \mathcal{A}_{-1} =
    \begin{bmatrix}
    0 & 0 & 0 \\
    0 & -4 & -2\sqrt{2} \\
    0 & -2\sqrt{2} & -2
    \end{bmatrix},
    \quad
    \mathcal{A}_{0} =
    \begin{bmatrix}
    0 & 0 & 0 \\
    0 & 0 & 0 \\
    0 & 0 & 0
    \end{bmatrix},
    \quad
    \mathcal{A}_{1} =
    \begin{bmatrix}
    0 & 0 & 0 \\
    0 & 4 & 2\sqrt{2} \\
    0 & 2\sqrt{2} & 2
    \end{bmatrix}.
\end{align}
Thus, $\text{Tr}[-\hat{S}_{z}\mathbb{G}_{0}^{\mathsf{A}}\mathbb{T}^{\mathsf{A}}]$ will scale like $R^{0 + 3} = R^{3}$ for small $R$. Interestingly, repeating the same work for $\hat{L}_{z}\mathbb{G}_{0}^{\mathsf{A}}$ (say, by taking $\hat{S}_{z} \to \hat{J}_{z} - \hat{S}_{z} = \hat{L}_{z}$ in the above expressions to avoid recalculating $\hat{L}_{z}\mathbf{A}_{JM}^{(1,2,3)}$) we find that the corresponding $R^{0}$ terms in $\hat{L}_{z}\mathbb{G}_{0}^{\mathsf{A}}$ vanish exactly (the matrices analogous to $\mathcal{A}_{1},\mathcal{A}_{0}, \mathcal{A}_{-1}$ are all zero), regardless of what $\mathbb{T}$ is. Of course, this is to be expected as the sum of the $\hat{S}_{z}$ and $\hat{L}_{z}$ terms in the small $R$ limit should result in the small $R$ limit of $\hat{J}_{z}$ term.

In sum, at least on a mathematical level the spin operator is the relevant operator in the definition of $\Phi_{J}$ for small $R/\lambda.$ One remark, however, is that spin and orbital contributions to $\Phi_{J}$ are
\begin{align}
    \Phi_{L} &=
    \frac{2}{\pi}\text{Tr}
    [(-\vecop{L}\mathbb{G}_{0})^{\mathsf{A}}(\mathbb{T}_{body}^{\mathsf{A}} - \mathbb{T}_{body}\mathbb{G}_{0}^{\mathsf{A}}\mathbb{T}_{body}^{\dagger})],
    \\
    \Phi_{S} &=
    \frac{2}{\pi}\text{Tr}
    [(-\vecop{S}\mathbb{G}_{0})^{\mathsf{A}}(\mathbb{T}_{body}^{\mathsf{A}} - \mathbb{T}_{body}\mathbb{G}_{0}^{\mathsf{A}}\mathbb{T}_{body}^{\dagger})
    ],
\end{align}
where, for example,
\begin{align}
    (\hat{S}_{a}\mathbb{G}_{0})^{\mathsf{A}}_{ij}(\mathbf{x}, \mathbf{y})
    = \frac{1}{2i}[\hat{S}_{a,ib} \mathbb{G}_{0,bj}(\mathbf{x}, \mathbf{y}) - (\hat{S}_{a,jb} \mathbb{G}_{0,bi}(\mathbf{y}, \mathbf{x}))^{*}]
    .
\end{align}
Namely, it is $(-\vecop{L}\mathbb{G}_{0})^{\mathsf{A}}$ and $(-\vecop{S}\mathbb{G}_{0})^{\mathsf{A}}$ rather
than $(-\vecop{L}\mathbb{G}_{0}^{\mathsf{A}})$ and $(-\vecop{S}\mathbb{G}_{0}^{\mathsf{A}})$ that originally
appear in what are deemed the orbital and spin contributions (see Eq.~\eqref{eq:pretorquetracesinglebody}) and
that are individually Hermitian. Since the total angular momentum operator commutes with
$\mathbb{G}_{0},$ the $(-\vecop{J}\mathbb{G}_{0})^{\mathsf{A}}$ term in the trace expression in $\Phi_{J}$
can be replaced with $(-\vecop{J}\mathbb{G}_{0}^{\mathsf{A}})$ which leads to more convenient analysis due
to the lower rank of $\mathbb{G}_{0}^{\mathsf{A}}$ compared to $\mathbb{G}_{0}.$ In general, $\vecop{L}$ and $\vecop{S}$ do not commute with $\mathbb{G}_{0}$ so this similar switch of, for example, $(-\vecop{L}\mathbb{G}_{0})^{\mathsf{A}} \to -\vecop{L}\mathbb{G}_{0}^{\mathsf{A}}$ is not correct. It is interesting to see
that in the small $R$ limit of the simplified $\Phi_{J}$ expression with $(-\hat{J}_{z}\mathbb{G}_{0}^{\mathsf{A}}),$ one can make the replacement $\hat{J}_{z}\to \hat{S}_{z}.$ However, as it is the total
angular momentum that is the assumed conserved quantity, physically it is likely that only
total angular momentum transfer is a meaningful quantity to calculate, and it is only in the strict point-particle limit that the replacement of $\hat{J}_{z}$ with $\hat{S}_{z}$ is exact; $(\hat{S}_{z}\mathbb{G}_{0})^{\mathsf{A}}(\mathbf{x}, \mathbf{y})$ and
$(\hat{L}_{z}\mathbb{G}_{0})^{\mathsf{A}}(\mathbf{x}, \mathbf{y})$ both appear to diverge as $\mathbf{y}
\rightarrow \mathbf{x},$ so it is not clear if calculating a separate quantity to designate as orbital and
spin is physically meaningful, in particular in the case of a single isolated finite-size object. This is reminiscent of
diverging energies in the Casimir force calculations~\cite{casimir1948attraction}, although the forces (related to the gradients of
the energies) are finite. Torque is a physically meaningful quantity, which solely depends on the total angular momentum transfer, and $(\hat{J}_{z}\mathbb{G}_{0})^{\mathsf{A}}_{ij}(\mathbf{x}, \mathbf{y}) = \hat{J}_{z}\mathbb{G}_{0,ij}^{\mathsf{A}}(\mathbf{x}, \mathbf{y})$ is manifestly free of singularities as $\mathbf{y} \rightarrow \mathbf{x}$, as can be seen from
$ \hat{J}_{z}\mathbb{G}_{0}^{\mathsf{A}}(\mathbf{x}, \mathbf{y}) =
  k\sum_{j,m}(-1)^{m} m\hbar [
    \mathbf{RM}_{j,m}(k\mathbf{x})\mathbf{RM}_{j,-m}(k\mathbf{y}) + \mathbf{RN}_{j,m}(k\mathbf{x})\mathbf{RN}_{j,-m}(k\mathbf{y})]$
or Eq.~\eqref{eq:JzG0Arr}. 
\end{widetext}

\section{$\hat{J}_{z}\mathbb{G}_{0}$ and $\hat{J}_{z}\mathbb{G}_{0}^{\mathsf{A}}$ Dyadic forms}

The Green's function dyadic can be written as
\begin{align}
    \mathbb{G}_{0} = \frac{e^{ikr}}{4\pi r}\left[a\mathbb{I} + b \mathbf{e}_{r}\otimes\mathbf{e}_{r}\right]
\end{align}
where $k = \omega/c,$ 
\begin{align}
    a &= 1 + \frac{ikr - 1}{(kr)^{2}}
\end{align}
\begin{align}
    b &= \frac{3 - 3ikr - (kr)^{2}}{(kr)^{2}}
\end{align}
and $\mathbf{e}_{r} = (\mathbf{R} - \mathbf{R}')/|\mathbf{R} - \mathbf{R}'|$ in a unit vector from the source location $\mathbf{R}'$ to the observation point $\mathbf{R}.$ Without loss of generality, we can consider the source location $\mathbf{R}'$ to be located at the origin of the coordinate system so that the relevant vectors can be expressed using $\mathbf{e}_{r}, \mathbf{e}_{\theta}, \mathbf{e}_{\phi}$ polar coordinate basis. \begin{widetext}

Working in the $\{\mathbf{e}_{r}, \mathbf{e}_{\theta}, \mathbf{e}_{\phi}\}$ basis, we have

\begin{align}
    \hat{S}_{z}\mathbb{G}_{0}(\mathbf{r}, 0) &=
    \frac{e^{ikr}}{4\pi r}\begin{bmatrix}
    0 & 0 & -i\sin{\theta}a \\
    0 & 0 & -i\cos{\theta}a \\
    i\sin{\theta}(a+b) & i\cos{\theta}a & 0
    \end{bmatrix} \\
    &=
    \frac{e^{ikr}}{4\pi r}\frac{1}{(kr)^{2}}\begin{bmatrix}
    0 & 0 & i\sin{\theta} \\
    0 & 0 & i\cos{\theta} \\
    2i\sin{\theta} & -i\cos{\theta} & 0
    \end{bmatrix} \nonumber \\
    & + \frac{e^{ikr}}{4\pi r}\frac{1}{kr}\begin{bmatrix}
    0 & 0 & \sin{\theta} \\
    0 & 0 & \cos{\theta} \\
    2\sin{\theta} & -\cos{\theta} & 0
    \end{bmatrix} \nonumber \\
    & + \frac{e^{ikr}}{4\pi r}\begin{bmatrix}
    0 & 0 & -i\sin{\theta} \\
    0 & 0 & -i\cos{\theta} \\
    0 & i\cos{\theta} & 0
    \end{bmatrix} \nonumber
\end{align}
Using $\frac{\partial}{\partial \phi} \mathbf{e}_{r} = \sin{\theta}\mathbf{e}_{r},$ we also find
\begin{align}
    \hat{L}_{z}\mathbb{G}_{0}(\mathbf{r}, 0) &=
    \frac{e^{ikr}}{4\pi r}
    \begin{bmatrix}
    0 & 0 & -ib\sin{\theta} \\
    0 & 0 & 0 \\
    -ib\sin{\theta} & 0 & 0
    \end{bmatrix} \\
    &=
    \frac{e^{ikr}}{4\pi r}\frac{1}{(kr)^{2}}\begin{bmatrix}
    0 & 0 & -3i\sin{\theta} \\
    0 & 0 & 0 \\
    -3i\sin{\theta} & 0 & 0
    \end{bmatrix} \nonumber \\
    & + \frac{e^{ikr}}{4\pi r}\frac{1}{kr}\begin{bmatrix}
    0 & 0 & -3\sin{\theta} \\
    0 & 0 & 0 \\
    -3\sin{\theta} & 0 & 0
    \end{bmatrix} \nonumber \\
    & + \frac{e^{ikr}}{4\pi r}\begin{bmatrix}
    0 & 0 & i\sin{\theta} \\
    0 & 0 & 0 \\
    i\sin{\theta} & 0 & 0
    \end{bmatrix} \nonumber
\end{align}
and, hence,
\begin{align}
    \hat{J}_{z}\mathbb{G}_{0}(\mathbf{r}, 0) &=
    \frac{e^{ikr}}{4\pi r}\begin{bmatrix}
    0 & 0 & -i\sin{\theta}(a+b) \\
    0 & 0 & -i\cos{\theta}a \\
    i\sin{\theta}a & i\cos{\theta}a & 0
    \end{bmatrix} \\
    &=
    \frac{e^{ikr}}{4\pi r}\frac{1}{(kr)^{2}}\begin{bmatrix}
    0 & 0 & -2i\sin{\theta} \\
    0 & 0 & i\cos{\theta} \\
    -i\sin{\theta} & -i\cos{\theta} & 0
    \end{bmatrix} \nonumber \\
    & + \frac{e^{ikr}}{4\pi r}\frac{1}{kr}\begin{bmatrix}
    0 & 0 & -2\sin{\theta} \\
    0 & 0 & \cos{\theta} \\
    -\sin{\theta} & -\cos{\theta} & 0
    \end{bmatrix} \nonumber \\
    & + \frac{e^{ikr}}{4\pi r}\begin{bmatrix}
    0 & 0 & 0 \\
    0 & 0 & -i\cos{\theta} \\
    i\sin{\theta} & i\cos{\theta} & 0
    \end{bmatrix}. \nonumber
\end{align}
In the Cartesian basis, this is
\begin{align}
    \hat{J}_{z}\mathbb{G}_{0}(\mathbf{r}, 0) &=
    \frac{e^{ikr}}{4\pi r}\bigg(
    a\begin{bmatrix}
    0 & -i & 0 \\
    i & 0 & 0 \\
    0 & 0 & 0
    \end{bmatrix}_{cart}
    +
    b
    \begin{bmatrix}
    i\cos{\phi}\sin{\theta}^{2}\sin{\phi} & -i\cos{\phi}^{2}\sin{\theta}^{2} & 0 \\
    i\sin{\theta}^{2}\sin{\phi}^{2} & -i\cos{\phi}\sin{\theta}^{2}\sin{\phi} & 0 \\
    i\cos{\theta}\sin{\theta}\sin{\phi} & -i\cos{\theta}\cos{\phi}\sin{\theta} & 0
    \end{bmatrix}_{cart}
    \bigg)
\end{align}
In this notation, in the $\{\mathbf{e}_{r}, \mathbf{e}_{\theta}, \mathbf{e}_{\phi}\}$ basis $\mathbb{G}_{0}^{\mathsf{A}}$ is
\begin{align}
    \mathbb{G}_{0}^{\mathsf{A}}(\mathbf{r}, 0) &=
    \frac{k}{4\pi }(\frac{\cos{kr}}{(kr)^{2}} - \frac{\sin{kr}}{(kr)^{3}} + \frac{\sin{kr}}{kr})\begin{bmatrix}
    1 & 0 & 0 \\
    0 & 1 & 0 \\
    0 & 0 & 1
    \end{bmatrix} \\
    & + 
    \frac{k}{4\pi }(-3\frac{\cos{kr}}{(kr)^{2}} +3 \frac{\sin{kr}}{(kr)^{3}} - \frac{\sin{kr}}{kr})\begin{bmatrix}
    1 & 0 & 0 \\
    0 & 0 & 0 \\
    0 & 0 & 0
    \end{bmatrix} \nonumber
\end{align}
and
\begin{align}
    \hat{J}_{z}\mathbb{G}_{0}^{\mathsf{A}}(\mathbf{r}, 0) &=
    \frac{k}{4\pi }(\frac{\cos{kr}}{(kr)^{2}} - \frac{\sin{kr}}{(kr)^{3}} + \frac{\sin{kr}}{kr})\begin{bmatrix}
    0 & 0 & -i\sin{\theta} \\
    0 & 0 & -i\cos{\theta} \\
    i\sin{\theta} & i\cos{\theta} & 0
    \end{bmatrix} \\
    & + 
    \frac{k}{4\pi }(-3\frac{\cos{kr}}{(kr)^{2}} +3 \frac{\sin{kr}}{(kr)^{3}} - \frac{\sin{kr}}{kr})\begin{bmatrix}
    0 & 0 & -i\sin{\theta} \\
    0 & 0 & 0 \\
    0 & 0 & 0
    \end{bmatrix} \nonumber \\
    &= 
    \frac{k}{4\pi }(\frac{\cos{kr}}{(kr)^{2}} - \frac{\sin{kr}}{(kr)^{3}} + \frac{\sin{kr}}{kr})\begin{bmatrix}
    0 & 0 & 0 \\
    0 & 0 & -i\cos{\theta} \\
    i\sin{\theta} & i\cos{\theta} & 0
    \end{bmatrix} \\
    & + 
    \frac{k}{4\pi }2\frac{\sin{kr}-kr\cos{kr}}{(kr)^{3}}\begin{bmatrix}
    0 & 0 & -i\sin{\theta} \\
    0 & 0 & 0 \\
    0 & 0 & 0
    \end{bmatrix} \nonumber
\end{align}
Changing from the $\{\mathbf{e}_{r}, \mathbf{e}_{\theta}, \mathbf{e}_{\phi}\}$ basis to the Cartesian basis $\{\mathbf{e}_{x}, \mathbf{e}_{y}, \mathbf{e}_{z}\}$ one finds
\begin{align}
    \hat{J}_{z}\mathbb{G}_{0}^{\mathsf{A}}(\mathbf{r}, 0) &=
    \frac{k}{4\pi }(\frac{\cos{kr}}{(kr)^{2}} - \frac{\sin{kr}}{(kr)^{3}} + \frac{\sin{kr}}{kr})
    \begin{bmatrix}
    0 & -i & 0 \\
    i & 0 & 0 \\
    0 & 0 & 0
    \end{bmatrix}_{cart} \\
    & + 
    \frac{k}{4\pi }(-3\frac{\cos{kr}}{(kr)^{2}} +3 \frac{\sin{kr}}{(kr)^{3}} - \frac{\sin{kr}}{kr})
    \begin{bmatrix}
    i\cos{\phi}\sin{\theta}^{2}\sin{\phi} & -i\cos{\phi}^{2}\sin{\theta}^{2} & 0 \\
    i\sin{\theta}^{2}\sin{\phi}^{2} & -i\cos{\phi}\sin{\theta}^{2}\sin{\phi} & 0 \\
    i\cos{\theta}\sin{\theta}\sin{\phi} & -i\cos{\theta}\cos{\phi}\sin{\theta} & 0
    \end{bmatrix}_{cart} \nonumber
\end{align}
The small $r$ expansions are
\begin{align}
    \mathbb{G}_{0}^{\mathsf{A}}(\mathbf{r}, 0)
    &=
    \frac{k}{6\pi}
    \begin{bmatrix}
    1 & 0 & 0 \\
    0 & 1 & 0 \\
    0 & 0 & 1
    \end{bmatrix}_{cart} + \mathcal{O}(kr)^{2} \\
    \hat{J}_{z}\mathbb{G}_{0}^{\mathsf{A}}(\mathbf{r}, 0)
    &=
    \frac{k}{6\pi}
    \begin{bmatrix}
    0 & -i & 0 \\
    i & 0 & 0 \\
    0 & 0 & 0
    \end{bmatrix}_{cart}
    + \mathcal{O}(kr)^{2} \label{eq:JzG0Arr} \\
    &= \frac{k}{6\pi}\hat{S}_{z} + \mathcal{O}(kr)^{2} \nonumber \\
    &= \hat{S}_{z}\mathbb{G}_{0}^{\mathsf{A}}(\mathbf{r}, 0) + \mathcal{O}(kr)^{2}
    \nonumber
\end{align}
Once again, we see that as the two spatial arguments approach one another, the total angular momentum operator in $\hat{J}_{z}\mathbb{G}_{0}^{\mathsf{A}}$ can be replaced with the spin operator $\hat{S}_{z}$ to leading order in the expansion. The replacement is exact only if the two spatial coordinates coincide. Insert factors of $\hbar$ in the intermediate expressions if working in units where $\hbar\neq 1$ in $\hat{J}_{z}, \hat{L}_{z}, \hat{S}_{z}.$
\end{widetext}

\section{Upper Bounds on Torque}
\label{app:torquebounds}
In this section we provide details of the calculation of bounds on maximal torque using techniques developed in Refs.~\cite{T_operator_bounds_angle_integrated,global_T_operator_bounds,molesky2020hierarchical,molesky2021comm} and reviewed in Ref.~\cite{chao2022physical}. Formally, the problem we solve is the maximization of $\Phi_{J}$ for an object contained within a spherical design domain $\Omega$ subject to the conservation of global resistive and reactive power:
\begin{align}
 &\max_{\{\left|\mathbf{T}_{n}\right>\in\Omega\}}~
 -\frac{2}{\pi}\sum_{n}\rho_{n}\left(\Im\left[
 \left<\mathbf{Q}_{n}|\hat{J}_{z}|\mathbf{T}_{n}\right>\right] - 
 \left<\mathbf{T}_{n}\right|
 \hat{J}_{z}\mathbb{G}_{0}^{\mathsf{A}}\left|\mathbf{T}_{n}\right>
 \right)
 \nonumber \\
 &\text{such that}~\forall n
 \nonumber \\
 &\Im\left[\left<\mathbf{Q}_{n}|\mathbf{T}_{n}\right>\right] - 
 \left<\mathbf{T}_{n}\right|
 \left(\mathbb{V}^{-1\dagger}
 -\mathbb{G}_{0}^{\dagger}\right)^{\mathsf{A}}
 \left|\mathbf{T}_{n}\right> = 0, \nonumber \\
  &\Re\left[\left<\mathbf{Q}_{n}|\mathbf{T}_{n}\right>\right] - 
 \left<\mathbf{T}_{n}\right|
 \left(\mathbb{V}^{-1\dagger}
 -\mathbb{G}_{0}^{\dagger}\right)^{\mathsf{S}}
 \left|\mathbf{T}_{n}\right> = 0.
 \label{totalPowerProb}
\end{align}
The induced currents $\mathbb{T}|\mathbf{Q}_{n}\rangle$ are taken as the optimization degrees of freedom. Here $\mathbb{G}_{0}^{\mathsf{S}} \equiv \frac{1}{2}(\mathbb{G}_{0} + \mathbb{G}_{0}^{\dagger})$ is the symmetric part of $\mathbb{G}_{0}.$
Global power conservation here means when spatially integrated over the object. There can still be local violations of power conservation. The constraints follow by acting on the left and right of $\mathbb{T} = \mathbb{T}^{\dagger}\big(\mathbb{V}^{-1\dagger} - \mathbb{G}_{0}^{\dagger}\big)\mathbb{T}$ by the eigenvectors of $\mathbb{G}_{0}^{\mathsf{A}},$ and are statements about conservation of energy (optical theorem~\cite{jackson1999classical}).

If only resistive power conservation is globally conserved (only the Im constraint is kept), the optimal value can be bound by a semi-analytical expression using Lagrange duality. Applying the relaxation of Lagrange duality (\cite{boyd2004convex,beck2006strong,angeris2019computational,
molesky2020t,angeris2021heuristic}) one gets the semianalytic expression presented in the main text (details of the calculation are given below).

Shown in Fig.~\ref{fig:torque_bounds} is also $\Phi_{J, opt}$ when reactive power conservation is imposed in addition to resistive power conservation (dashed lines). It is seen that including reactive power conservation can lead to substantially tighter limits, particularly for dielectric materials and smaller design domains. The solution to Eq.~\eqref{totalPowerProb} with resistive and reactive power conservation does not, in general, have a simple semi-analytic expression similar to Eq.~\eqref{realPowerSol} with only resistive power conservation.
The optimization problem was solved numerically using a modification of the code developed and provided in Ref.~\cite{molesky2020t}. For large $R/\lambda$, the bounds suggest that the response is mostly dominated by conservation of resistive power as there is enough design freedom in the optimization problem to satisfy resonance conditions, so the inclusion of reactive power conservation does not lead to substantial tightening of bounds.

Modifying the fluctuating-volume current formulation and codes developed in Refs.~\cite{polimeridis2014computation,polimeridis2015fluctuating,polimeridisgithub}, from power quantities to the torque quantities derived in this article, we discovered design patterns that approach the torque bounds. Constraints to make the design pattern experimentally practical to fabricate were ignored, for simplicity. 
Shown in Fig.~\ref{fig:torque_bounds} as an inset is a sample structure for $R/\lambda = 0.25$ and $\chi = -10 + i.$
Intuitively, one expects a chiral object to be an optimal performer when the electric susceptibility is isotropic (without anisotropy, there is only geometrical structure freedom left with which to discriminate incoming waves).
Indeed, the optimal induced current $\left|\mathbf{J}^{(opt)}\right> = -\frac{i}{kZ}\left|\mathbf{T}^{(opt)}\right>$  is a sum of terms with only one sign of $m$, in agreement with intuition that the structure of an optimal body is such that the induced currents are biased towards one azimuthal direction.
Some of the chiral structures from the inverse designs approach the bounds within a factor of 15. Adding local power conservation constraints may result in better agreement~\cite{molesky2020hierarchical,kuang2020computational}.

\subsection{Semi-analytical bounds for spherical bounding domains}

Here we explain in more detail the derivation of the bounds when only imposing the asymmetric constraints (physically, the conservation of resistive power) over the entire domain (a spherical ball of radius $R$). We calculate a bound on the optimization problem by calculating the Lagrange dual function~\cite{boyd2004convex}. The corresponding Lagrangian that involves $(t,j,m)$ terms is 
\begin{widetext}
\begin{align}
    \mathcal{L}^{(t,j,m)} = -m\rho_{t,j,m}
    \textrm{Im}[\braket{\mathbf{S}_{t,j,m}}{\mathbf{T}_{t,j,m}}] 
    + m\rho_{t,j,m}\bra{\mathbf{T}_{t,j,m}}\mathbb{G}_{0}^{\mathsf{A}} \ket{\mathbf{T}_{t,j,m}}) \nonumber \\
    +\alpha(\textrm{Im}[\braket{\mathbf{S}_{t,j,m}}{\mathbf{T}_{t,j,m}}] - \bra{\mathbf{T}_{t,j,m}}\mathbb{U}^{\mathsf{A}}\ket{\mathbf{T}_{t,j,m}}).
\end{align}
\end{widetext}
Here, $\mathbb{U} \equiv \mathbb{V}^{-1\dagger} - \mathbb{G}_{0}^{\dagger}.$ The main observation is the following. The optimal $\ket{\mathbf{T}_{t,j,m}}$ at a stationary point satisfies
\begin{align}
    (-m\rho_{t,j,m}\mathbb{G}_{0}^{\mathsf{A}} +  \alpha\mathbb{U}^{\mathsf{A}})\ket{\mathbf{T}_{t,j,m}} \nonumber\\
    = \left(-\frac{m\rho_{t,j,m} i}{2} + \frac{i\alpha}{2}\right)\ket{\mathbf{S}_{t,j,m}}.
    \label{eq:stationarypointeq}
\end{align}
Define $\zeta = k^{2}\lVert\chi\rVert^{2}/\Im\left[\chi\right].$ In general,
\begin{align}
    \ket{\mathbf{T}_{t,j,m}} = \frac{i}{2}
    \frac{(-m\rho_{t,j,m} + \alpha)}{-m\rho_{t,j,m}^{2} + \alpha(\frac{1}{\zeta} + \rho_{t,j,m})}
    \ket{\mathbf{S}_{t,j,m}} + \ket{\mathbf{K}},
    \label{eq:Topt}
\end{align}
where $\ket{\mathbf{K}}$ lies in the kernel of the operator which multiplied $\ket{\mathbf{T}_{t,j,m}}$ in Eq.~\eqref{eq:stationarypointeq}.  This may be used to derive a semi-analytical expression for the optimal dual objective value. There are a few cases to consider.

\begin{itemize}
    \item If $m = 0,$ then the contribution to the torque is clearly 0.
    \item If  $m > 0,$ there are 3 cases for $\alpha$ to consider.
    \begin{itemize}
        \item If $\alpha >0,$ then the dual function is unbounded since 
            $(-m\rho_{t,j,m}\mathbb{G}_{0}^{\mathsf{A}} +  \alpha\mathbb{U}^{\mathsf{A}})$
        becomes indefinite (recall that $\mathbb{G}_{0}^{\mathsf{A}}$ is positive semidefinite).
        \item If $\alpha = 0,$ then the operator is negative semidefinite. In such a case,
            \begin{align}
                \ket{\mathbf{T}_{t,j,m}} = \frac{i}{2}
            \frac{1}{\rho_{t,j,m} }
            \ket{\mathbf{S}_{t,j,m}} + \ket{\mathbf{K}}
            \end{align}
            where $\ket{\mathbf{K}}$ is in the kernel of $-m\rho_{t,j,m}\mathbb{G}_{0}^{\mathsf{A}}.$ Evaluating the objective function for this vector gives $-\frac{m}{4}\braket{\mathbf{S}_{t,j,m}}{\mathbf{S}_{t,j,m}}.$
        \item If $\alpha < 0,$ then (since $\mathbb{U}^{\mathsf{A}}$ is positive definite) the kernel is trivial. Then $\alpha$ can be solved for from
            \begin{align}
                \textrm{Im}[\braket{\mathbf{S}_{t,j,m}}{\mathbf{T}_{t,j,m}}] = \bra{\mathbf{T}_{t,j,m}}\mathbb{U}^{\mathsf{A}}\ket{\mathbf{T}_{t,j,m}}.
            \end{align}
            The two solutions are $\alpha = m\rho_{t,j,m}$ and $\alpha = m\rho_{t,j,m}(2\rho_{t,j,m} - u_{t,j,m})/u_{t,j,m},$ where $u_{t,j,m} = \frac{1}{\zeta} + \rho_{t,j,m}.$ The maximum of the objective evaluated at these two values of $\alpha$ is given by
            \begin{align}
                \text{max}\left(0, -m\left(\frac{\rho_{t,j,m}}{u_{t,j,m}} - \left(\frac{\rho_{t,j,m}}{u_{t,j,m}}\right)^{2} \right)  \right).
            \end{align}
            Note that $\alpha < 0$ when $\frac{\rho_{t,j,m}}{u_{t,j,m}} < \frac{1}{2}.$ It hits $\alpha = 0^{-}$ when $\frac{\rho_{t,j,m}}{u_{t,j,m}} = \frac{1}{2}.$
        \item In sum, for $m > 0$ the optimal objective value is
        \begin{align}
            \mathcal{L}_{opt}^{(t,j,m)} =
            \begin{cases}
            \text{max}\left(0, -\frac{m}{4} \right), \text{ if } \frac{\rho_{t,j,m}}{u_{t,j,m}} \geq \frac{1}{2}\\
            \text{max}\left(0, -m\left(\frac{\rho_{t,j,m}}{u_{t,j,m}} - \left(\frac{\rho_{t,j,m}}{u_{t,j,m}}\right)^{2} \right) \right), \text{ if } \frac{\rho_{t,j,m}}{u_{t,j,m}} < \frac{1}{2}\\
            \end{cases}
        \end{align}
        This simplifies for $m > 0$ as the above is always 0. That is, the positive $m$ vector spherical harmonics do not contribute to the objective at the optimal solution. Intuitively, only one sign should contribute to the torque if one wishes to maximize the torque imparted to an object.
    \end{itemize}
    
    \item If $m < 0,$ there are three cases of $\alpha$ to consider.
        \begin{itemize}
            \item If $\alpha > 0,$ then $-m\rho_{t,j,m}\mathbb{G}_{0}^{\mathsf{A}} + \alpha\mathbb{U}^{\mathsf{A}}$ is positive definite, so the kernel is trivial. Then $\alpha$ can be solved for from
            \begin{align}
                \textrm{Im}[\braket{\mathbf{S}_{t,j,m}}{\mathbf{T}_{t,j,m}}] = \bra{\mathbf{T}_{t,j,m}}\mathbb{U}^{\mathsf{A}}\ket{\mathbf{T}_{t,j,m}}.
            \end{align}
            The two solutions are $\alpha = m\rho_{t,j,m}$ and $\alpha = m\rho_{t,j,m}(2\rho_{t,j,m} - u_{t,j,m})/u_{t,j,m},$ where $u_{t,j,m} = \frac{1}{\zeta} + \rho_{t,j,m}.$ The maximum of the objective evaluated at these two values of $\alpha$ is given by
            \begin{align}
                \text{max}\left(0, -m\left(\frac{\rho_{t,j,m}}{u_{t,j,m}} - \left(\frac{\rho_{t,j,m}}{u_{t,j,m}}\right)^{2} \right)  \right).
            \end{align}
            Note that $\alpha > 0$ when $\frac{\rho_{t,j,m}}{u_{t,j,m}} < \frac{1}{2}.$ It hits $\alpha = 0^{+}$ at $\frac{\rho_{t,j,m}}{u_{t,j,m}} = \frac{1}{2}.$
            \item If $\alpha = 0,$ then the operator is positive semi-definite. In such a case,
            \begin{align}
                \ket{\mathbf{T}_{t,j,m}} = \frac{i}{2}
            \frac{1}{\rho_{t,j,m} }
            \ket{\mathbf{S}_{t,j,m}} + \ket{\mathbf{K}}
            \end{align}
            where $\ket{\mathbf{K}}$ is in the kernel of $-m\rho_{t,j,m}\mathbb{G}_{0}^{\mathsf{A}}.$ Evaluating the objective function for this vector gives $-\frac{m}{4}\braket{\mathbf{S}_{t,j,m}}{\mathbf{S}_{t,j,m}}.$
            \item If $\alpha < 0,$ then the operator is indefinite and the dual function diverges.
            \item In sum, for $m < 0$ the optimal objective value is
            \begin{align}
            \mathcal{L}_{opt}^{(t,j,m)} =
            \begin{cases}
            \text{max}\left(0, -\frac{m}{4} \right), \text{ if } \frac{\rho_{t,j,m}}{u_{t,j,m}} \geq \frac{1}{2}\\
            \text{max}\left(0, -m\left(\frac{\rho}{u} - \left(\frac{\rho}{u}\right)^{2} \right) \right), \text{ if } \frac{\rho_{t,j,m}}{u_{t,j,m}} < \frac{1}{2}\\
            \end{cases}
        \end{align}
        \end{itemize}

\end{itemize}
In sum, simplifying the calculations, the semi-analytical result for the $(t,j,m)$ block is given by
\begin{widetext}
    \begin{align}
        \mathcal{L}_{opt}^{(t,j,m)} &=
        \begin{cases}
        0, &\text{ if } m \geq 0 \\
        \begin{cases}
                \text{max}\left(0, -\frac{m}{4} \right) &\text{ if } \frac{\rho_{t,j,m}}{u_{t,j,m}} \geq \frac{1}{2}\\
                \text{max}\left(0, -m\left(\frac{\rho_{t,j,m}}{u_{t,j,m}} - \left(\frac{\rho_{t,j,m}}{u_{t,j,m}}\right)^{2} \right) \right) &\text{ if } \frac{\rho_{t,j,m}}{u_{t,j,m}} < \frac{1}{2}\\
        \end{cases}
        &\text{ if } m < 0.
        \end{cases}
    \end{align}
    which can be written in terms of $\zeta$ and $\rho_{t,j,m}$ as
    \begin{align}
        \mathcal{L}_{opt}^{(t,j,m)} &=
        \begin{cases}
        0, &\text{ if } m \geq 0 \\
        \begin{cases}
                -\frac{m}{4} &\text{ if } \zeta\rho_{t,j,m} \geq 1\\
                -\frac{m\zeta\rho_{t,j,m}}{(1 + \zeta\rho_{t,j,m})^{2}} &\text{ if } \zeta\rho_{t,j,m} < 1\\
        \end{cases}
        &\text{ if } m < 0.
        \end{cases}
    \end{align}
\end{widetext}
This semi-analytical expression for the bound when only imposing global resistive power conservation is compared in the main text to the bounds found numerically when imposing global resistive and reactive power conservation. Note that one can rescale the variables by $k^{2}$ by redefining $\zeta = \lVert\chi\rVert^{2}/\Im\left[\chi\right]$ and $\rho_{t,j,m}$ as the eigenvalues of $k^{2}\mathbb{G}_{0}^{\mathsf{A}},$ making $\zeta$ and the eigenvalues dimensionless.

\section{Torque expressions in the point-particle limit}

Using the point particle limit and the Born approximation, a simplified expression for the torque exerted on particle 1 by particle 2 is
\begin{align}
    \boldsymbol{\tau}_{2}^{(1)}(T)\cdot\mathbf{e}_{z}
    &=
    -\frac{2}{\pi}\int_{0}^{\infty}d\omega n(\omega, T)
    \text{Im}\text{Tr}
    [
    \hat{J}_{z}\mathbb{G}_{0}\mathbb{T}_{2}^{\mathsf{A}}\mathbb{G}_{0}^{\dagger}\mathbb{T}_{1}^{\dagger}
    ]
\end{align}
Using the scattering operators in the small-sphere limit~\cite{asheichyk2017RHTPPs}, 
\begin{align}
    \text{Tr}
    [
    \hat{J}_{z}\mathbb{G}_{0}\mathbb{T}_{2}^{\mathsf{A}}\mathbb{G}_{0}^{\dagger}\mathbb{T}_{1}^{\dagger}
    ]
    =
    &\text{Im}\frac{9\omega^{4}}{c^{4}}\int_{V_{1}}d^{3}\mathbf{r}\int_{V_{2}}d^{3}\mathbf{r'}
    (\hat{J}_{z}\mathbb{G}_{0})_{ab}(\mathbf{r}, \mathbf{r}')
    \nonumber \\
    &\times\left(\frac{\MatrixVariable{\epsilon}_{2} - 1}{\MatrixVariable{\epsilon}_{2}+2}\right)_{bc}^{\mathsf{A}}
    \mathbb{G}_{0,cd}^ {\dagger}(\mathbf{r}', \mathbf{r})\left(\frac{\MatrixVariable{\epsilon}_{1} - 1}{\MatrixVariable{\epsilon}_{1}+2}\right)_{da}^{\dagger}.
\end{align}
Since the dimensions of the point particles are assumed small compared to any other dimensions in the problem, $(\hat{J}_{z}\mathbb{G}_{0})(\mathbf{r}, \mathbf{r}')$ and $\mathbb{G}_{0}^{\dagger}(\mathbf{r}', \mathbf{r})$ do not vary significantly between different points in the different particles. Letting $\mathbf{r}_{1}$ and $\mathbf{r}_{2}$ denote the centers of particle 1 and particle 2, respectively, the integrals $\int_{V_{1}}$ and $\int_{V_{2}}$ simply introduce factors of $4\pi R_{1}^{3}/3$ and $4\pi R_{2}^{3}/3$ so that the torque is proportional to the volumes of the particles. Compactly,
\begin{align}
    \boldsymbol{\tau}_{2}^{(1)}(T)\cdot\mathbf{e}_{z}
    = -\frac{2}{\pi}
    &\int_{0}^{\infty}d\omega \frac{\omega^{4}}{c^{4}} n(\omega,T) \times \nonumber \\
    &\text{Im}\text{Tr}_{cmp}[(\hat{J}_{z}\mathbb{G}_{0})(\mathbf{r}_{1}, \mathbf{r}_{2})\MatrixVariable{\alpha}^{\mathsf{A}}_{2}\mathbb{G}_{0}^{\dagger}(\mathbf{r}_{2}, \mathbf{r}_{1})\MatrixVariable{\alpha}_{1}^{\dagger}]
\end{align}
where $\text{Tr}_{cmp}$ means a trace only over the vector components ($\text{Tr}_{cmp}[\mathbb{A}(\mathbf{r}, \mathbf{r}')] \equiv \sum_{a} \mathbb{A}_{aa}(\mathbf{r}, \mathbf{r}')$).
Using very similar arguments, one finds
\begin{align}
    &\boldsymbol{\tau}_{1}^{(1)}(T)\cdot\mathbf{e}_{z}
    =
    -\frac{2}{\pi}\int_{0}^{\infty}d\omega n(\omega, T)\times
    \nonumber \\
    &\text{Im}\text{Tr}
    [
    \hat{J}_{z}\mathbb{G}_{0}\mathbb{T}_{1}^{\mathsf{A}}
    -
    \hat{J}_{z}\mathbb{G}_{0}\mathbb{T}_{1}\mathbb{G}_{0}^{\mathsf{A}}\mathbb{T}_{1}^{\dagger}
    +
    \hat{J}_{z}\mathbb{G}_{0}\mathbb{T}_{2}\mathbb{G}_{0}\mathbb{T}_{1}^{\mathsf{A}}
    ] \\
    &= -\frac{2}{\pi}\int_{0}^{\infty}d\omega \frac{\omega^{2}}{c^{2}} n(\omega,T) \times \nonumber \\
    &\Bigg(\text{Tr}_{cmp}
    [
    (\hat{J}_{z}\mathbb{G}_{0}^{\mathsf{A}})(\mathbf{r}_{1},\mathbf{r}_{1})\bigg(
    \MatrixVariable{\alpha}_{1}^{\mathsf{A}}
    -
    \frac{\omega^{2}}{c^{2}}\MatrixVariable{\alpha}_{1}\mathbb{G}_{0}^{\mathsf{A}}(\mathbf{r}_{1},\mathbf{r}_{1})\MatrixVariable{\alpha}_{1}^{\dagger}
    \bigg)
    ] \nonumber \\
    & +\frac{\omega^{2}}{c^{2}}\text{Im}\text{Tr}_{cmp}
    [
    (\hat{J}_{z}\mathbb{G}_{0})(\mathbf{r}_{1},\mathbf{r}_{2})
    \MatrixVariable{\alpha}_{2}\mathbb{G}_{0}(\mathbf{r}_{2},\mathbf{r}_{1})\MatrixVariable{\alpha}_{1}^{\mathsf{A}}
    ]
    \Bigg).
\end{align}
Similar arguments are used to get the torque expression in the single body case, Eq.~\eqref{eq:torquetracesinglePP}.

\section{InSb material parameters}

We consider particles with permittivities
\begin{align}
    \MatrixVariable{\epsilon} = 
    \begin{bmatrix}
    \epsilon_{1} & -i\epsilon_{2} & 0 \\
    i\epsilon_{2} & \epsilon_{1} & 0 \\
    0 & 0 & \epsilon_{3}
    \end{bmatrix}_{cart}.
\end{align}
For InSb one has~\cite{ott2019magnetothermoplasmonics}
\begin{align}
    \epsilon_{1} &= \epsilon_{\infty}
    \bigg(
    1 + \frac{\omega_{L}^{2} - \omega_{T}^{2}}{\omega_{T}^{2} - \omega^{2} - i\Gamma\omega} +
    \frac{\omega_{p}^{2}(\omega + i\gamma)}{\omega[\omega_{c}^{2} - (\omega + i\gamma)^{2}]}
    \bigg)\\
    \epsilon_{2} &= \frac{\epsilon_{\infty}\omega_{p}^{2}\omega_{c}}{\omega[(\omega + i\gamma)^{2} - \omega_{c}^{2}]} \\
    \epsilon_{3} &=
    \epsilon_{\infty}
    \bigg(
    1 + \frac{\omega_{L}^{2} - \omega_{T}^{2}}{\omega_{T}^{2} - \omega^{2} - i\Gamma\omega} -
    \frac{\omega_{p}^{2}}{\omega(\omega + i\gamma)}
    \bigg)
\end{align}
where $\epsilon_{\infty} = 15.7, \omega_{L} = 3.62\times 10^{13}$ rad/s, $\omega_{T} = 3.39\times 10^{13}$ rad/s, $n = 1.07\times 10^{17}$ cm$^{-3}, m^{*} = 1.99\times 10^{-32}$ kg, $\omega_{p} = \sqrt{\frac{nq^{2}}{m^{*}\epsilon_{0}\epsilon_{\infty}}} = 3.15\times 10^{13}$ rad/s, $q = 1.6\times 10^{-19} $ C, $\Gamma = 5.65 \times 10^{11}$ rad/s, $\gamma = 3.39\times 10^{12}$ rad/s, and $\omega_{c} = \frac{eB}{m^{*}}.$ To calculate the moment of inertia of the InSb particles, we used a density of $5.78$ g/cm$^{3}$.
\bibliography{refs}
\end{document}